\documentclass{article}
\usepackage[a4paper, total={16cm, 24.7cm}, twoside]{geometry}
\usepackage[utf8]{inputenc}
\usepackage{stfloats}
\usepackage{graphicx}
\usepackage{caption}
\usepackage{subcaption}
\usepackage{standalone}
\usepackage{amsmath}
\usepackage{amssymb}
\usepackage{booktabs}
\usepackage{diagbox}
\usepackage{pdfpages}
\usepackage{tabularx}
\usepackage{textcomp}
\usepackage{enumitem}
\usepackage{url}
\usepackage{todonotes}
\usepackage{lipsum}
\usepackage{hyperref}
\usepackage{url}
\usepackage{wrapfig}
\usepackage{float}
\usepackage{listings}
\usepackage{epstopdf}
\usepackage{tikz}
\usepackage{relsize}
\usepackage{mathtools}
\usepackage{cleveref}
\usepackage{latexsym}
\usepackage{blindtext}
\usepackage{graphics}
\usepackage{psfrag}   
\usepackage{listings}
\usepackage{fancyhdr}
\usepackage[printonlyused]{acronym} 
\usepackage{makeidx} 
\makeindex  
\usepackage{color,soul}
\usepackage{amsbsy}
\usepackage{lscape}
\usepackage{changepage}   
\usepackage[titletoc,toc,page]{appendix}
\usepackage{siunitx}
\usepackage[labelfont=bf]{caption}
\usepackage{enumerate}
\usepackage[super]{nth}
\usepackage{amsmath}
\usepackage{graphicx}
\usepackage{gensymb}
\usepackage{comment}
\usepackage{textcomp}
\usepackage{multirow}
\usepackage[section]{placeins}
\usepackage[official]{eurosym}
\usepackage[maxbibnames=99,sorting=none]{biblatex}
\usepackage{amsmath}
\usepackage{algorithm}
\usepackage{algpseudocode}
\usepackage{breqn}
\makeatletter
\renewcommand*\env@matrix[1][\arraystretch]{
  \edef\arraystretch{#1}
  \hskip -\arraycolsep
  \let\@ifnextchar\new@ifnextchar
  \array{*\c@MaxMatrixCols c}}
\makeatother

\title{Domain decomposition of large neural network surrogate models}

\author{T. G\"{o}dde $^{1,*}$, E. H. Atzema $^2$, B. Rosi\'{c} $^{1,3}$\\[0.8em]
\parbox{\textwidth}{
\raggedright
{\small \hangindent=1.8em $^1$ \quad University of Twente, Faculty of Engineering Technology, Applied Mechanics and Data Analysis, PO \\ \hspace{1.5em} Box 217, 7500AE, Enschede, the Netherlands
\\[0.1em]
$^2$ \quad \hangindent=1.8em Tata Steel Nederland Technology B.V., PO Box 10000, NL-1970 CA IJmuiden, the Netherlands
\\[0.1em]
$^3$ \quad \hangindent=1.8em Vienna University of Technology, E307-04-01 Digital Engineering group, Institute of Engineering Design and Product Development, Lehargasse 6, 1060 Vienna
\\[0.1em]
$^*$ \quad \hangindent=1.8em Correspondence: t.godde@utwente.nl}
}}

\begin{document}

\maketitle

\begin{abstract}
\noindent Neural networks (NNs) have gained significant attention across various engineering disciplines, particularly in design optimization, where they are commonly used to build surrogate models for high-dimensional regression problems. 
Despite their power as global approximators, NNs often fail to accurately capture local nonlinearities without relying on a large number of training parameters - resulting in slower training times.
To address these limitations, in this paper we propose domain decomposition methods (DDM), which divide the input feature space into multiple local subdomains, each modeled by a simpler NN, trained in parallel.
To recover the accuracy of a global approximation, interface constraints are introduced in the local loss functions to enforce continuity between subdomains.
The interface constraints are enforced with two different approaches: by utilizing Lagrange multipliers or augmented Lagrange multipliers methods. 
Both approaches are validated using synthetic data from a 2D linear-elastic compression problem, numerically solved using the finite element method.
The study investigates computational time and accuracy across varying numbers of subdomains to identify optimal partitioning strategies.
Compared to unconstrained approximations, both methods significantly improve continuity across subdomain interfaces.
Also, the use of DDMs improves approximation accuracy in nonlinear regions when compared to standard global NN training.
In terms of convergence, the augmented Lagrange method outperforms the standard Lagrange formulation by converging faster due to lower convergence requirements, albeit with a slightly lower accuracy.
However, its scalability makes it the preferred choice for large-scale problems, as the faster convergence outweighs the minor loss in accuracy.
Overall, these results highlight the augmented Lagrange method as a promising DDM approach for training efficient and scalable NN surrogate models.
\end{abstract}

\newcommand{\loss}{\mathcal{J}}
\newcommand{\Lagr}{\mathcal{L}}
\newcommand{\gebiet}{\mathcal{G}}
\newcommand{\gebietf}{\varOmega}
\newcommand{\ixa}{i}
\newcommand{\ixb}{j}
\newcommand{\ixc}{k}
\newcommand{\bound}{\varGamma_{\ixa\ixb}}
\newcommand{\boundi}{\varGamma_{\ixa\ixb}}
\newcommand{\uvec}{\boldsymbol u}
\newcommand{\point}{\boldsymbol x}
\newcommand{\wm}{\ws_{\ixa\ixb}}
\newcommand{\fm}{\fp_{\wm}}
\newcommand{\ls}{\boldsymbol{\lambda}}
\newcommand{\li}{\boldsymbol{\lambda}_\ixa}
\newcommand{\fli}{\boldsymbol{\lambda}^{(\iterli)}_\ixa}
\newcommand{\Constr}{\boldsymbol{\mathcal{Q}_i}}
\newcommand{\Cwi}{\Constr(\wi)}
\newcommand{\Cwii}{\Constr(\wii)}
\newcommand{\Cwiii}{\Constr(\wiii)}
\newcommand{\Cwj}{\bar{\Constr}(\wi)}
\newcommand{\Cwjj}{\bar{\Constr}({\wii})}
\newcommand{\Cwjjj}{\bar{\Constr}({\wiii})}
\newcommand{\Q}{\boldsymbol{\mathcal{Q}}}
\newcommand{\id}{\vek{1}}
\newcommand{\real}{\mathbb{R}}
\newcommand{\fvar}{\uvec}
\newcommand{\Fvar}{A}
\newcommand{\fp}{\hat{\fvar}}
\newcommand{\fvi}{\fvar_{\ixa}}
\newcommand{\fwi}{\fp_{\wi}}
\newcommand{\fwj}{\fp_{\wj}}
\newcommand{\fwk}{\fp_{\wk}}
\newcommand{\ws}{\boldsymbol{\theta}}
\newcommand{\wi}{\ws_{\ixa}}
\newcommand{\wii}{\ws_{\ixa}^{k}}
\newcommand{\wiii}{\ws_{\ixa}^{k+1}}
\newcommand{\wj}{\ws_{\ixb}}
\newcommand{\wk}{\ws_{\ixc}}
\newcommand{\w}{W}
\newcommand{\be}{b}
\newcommand{\xk}{\boldsymbol \point_\ixc}
\newcommand{\xoi}{\boldsymbol \point_\ixa}
\newcommand{\beq}{\begin{equation}}
\newcommand{\eeq}{\end{equation}}
\newcommand{\nat}{\mathbb{N}}
\newcommand{\pen}{\rho}
\newcommand{\al}{ALMA\xspace}
\newcommand{\nl}{LMA\xspace}
\newcommand{\n}{n}
\newcommand{\iter}{d}
\newcommand{\ncons}{N_{Q_i}}
\newcommand{\spatialvar}{x}
\newcommand{\stochasticvar}{\omega}
\newcommand{\vars}{(\spatialvar, \zeta(\stochasticvar))}
\newcommand{\normal}{\eta}
\newcommand{\pr}{p}
\newcommand{\du}{d}
\newcommand{\iterli}{\du}
\newcommand{\iterwi}{\pr}
\newcommand{\iterlin}{lin}
\newcommand{\mili}{N_{\du}}
\newcommand{\miwi}{N_{\pr}}
\newcommand{\milin}{N_{lin}}
\newcommand{\tolli}{\varepsilon^{\du}_{tol}}
\newcommand{\tolwi}{\varepsilon^{\pr}_{tol}}
\newcommand{\tollin}{\varepsilon^{lin}_{tol}}
\newcommand{\aleq}{\gets}
\section{Introduction}
A major challenge in computational mechanics-based design optimization lies in the efficient evaluation of spatio-temporal quantities of interest (QoIs), such as stress and strain fields, which are essential for assessing mechanical performance. These fields are typically computed using finite element method (FEM) solvers, rooted in the governing physics-based principles. However, in the context of design optimization, additional parametric dimensions are introduced through design variables, significantly increasing the problem's complexity and computational cost. These parameters are often described by corresponding probability distributions, giving rise to a stochastic optimization problem \cite{beyer2007}.
One approach for solving such problems is to perform a random walk in the input space—e.g., as proposed in \cite{Chan_2013}—where a new FEM simulation is computed online at each sampled design point to evaluate whether the resulting output improves the objective. However, this approach tends to converge slowly, as the random search frequently explores poor directions, leading to many discarded design evaluations and inefficient use of computational resources.
A more efficient approach involves first approximating the mapping between the set of design parameters and the QoI using a surrogate model \cite{forester08}, and then utilizing this surrogate to identify optimal design parameters. One classical surrogate technique is the polynomial chaos expansion \cite{wiener}, which can be constructed using a least-squares regression approach \cite{Zhou2015k}. However, in high-dimensional parametric spaces, polynomial chaos expansion often requires a prohibitively large number of FEM simulations due to the curse of dimension. To mitigate this issue, NN surrogates have been proposed as more sample-efficient alternatives \cite{forester08}. When equipped with appropriate architectures and activation functions, such networks can achieve high approximation accuracy with significantly fewer training samples, potentially exhibiting exponential gains in efficiency \cite{Barron1993}.
A drawback of NN surrogate models is that they rely on a global approximation, which interferes with precision requirements for the local spatio-temporal QoIs. 
To allow high precision, NNs are often highly over-parametrized, which leads to long training times, especially in high-dimensional settings. To keep the precision requirements and reduce the number of parameters, i.e., reduce training times, for high-dimensional NNs an alternative modeling approach is required.

DDMs \cite{ToselliWidlund2005} are capable of splitting global approximations into multiple connected local approximations by partitioning the input space.
Consequently, DDMs reduce parameters and training times of NNs by utilizing simpler, local approximations that can be computed in parallel.
Thus, instead of training one large NN, several local NNs are trained on local subdomains.
This is achieved by defining the NN weights locally, which reduces the number of parameters and enhances approximation accuracy.
These local NNs can be trained in parallel, significantly decreasing training times for large models.
However, the computational cost reduction comes at the expense of introducing additional unknowns to the problem - those describing the interfaces between local NNs. 
To handle interface conditions introduced by the decomposition, two main groups of DDMs are used, overlapping \cite{Schwarz,Lions} and non-overlapping, i.e., sub-structuring methods \cite{Lions2}.

Overlapping DDMs have emerged as an effective approach to improve the scalability of Physics-Informed Neural Networks (PINN) \cite{Raissi2017, KHH1} when solving high-dimensional differential equations.
PINNs are NNs in which the loss function is modified to solve boundary value problems, i.e., partial differential equations (PDE) with constraints. Specifically, the loss function is defined as a sum of mean squared error (sMSE) residuals of the problem's equations. 
One overlapping method is the deep domain decomposition method (D3M) \cite{KLi2019,WLi2020}, which solves the variational form a PDEs based on the deep Ritz method \cite{DRitz} with local sMSE residual loss functions. 
Another D3M \cite{WLi2020} uses the basic PINN formulation, i.e., strong formulation of a PDE and defines it as local sMSE residual loss function. 
Both methods enforce continuity along the interfaces by including these as additional sMSE penalty terms in the local loss functions.
Due to the inherent difficulty in balancing the additional MSE terms within the sMSE loss function, in another article the finite-basis PINN (FBPINN) \cite{Moseley} is defined by multiplying local NN approximations by window functions. 
These window functions constrain the output of each local network to its corresponding domain, ensuring that predictions vanish outside the local region. Continuity across domains is achieved implicitly, as overlapping NNs share their contributions through window functions.
As such, in an update, the local NN gradients are calculated from the global loss function.
However, determining the appropriate size of the overlap for convergence remains challenging. 
Simultaneously, the regions which are approximated by multiple local NNs introduce additional computational effort.
Despite their effectiveness, overlapping methods deal with additional computational cost due to overlap, which motivates the exploration of alternative decomposition strategies.

Non-overlapping DDMs improve computational efficiency by restricting communication between subdomains to the domain boundaries.
Two notable examples for sub-structuring DDMs for PINNs are extended PINN (XPINN) and conservative PINN (cPINN) \cite{Jagtap2020-1, Jagtap2020-2, Jang}. 
Similar to the overlapping methods, XPINNs and cPINNs employ additional sMSE penalty terms to enforce continuity across interfaces. 
Specifically, they constrain the function gradient and average of approximation losses along interfaces through penalty terms in the local sMSE loss functions. 
The difference between these two methods is that XPINNs are more flexible in choice of PDEs and interface conditions.
A recent PINN DMM has been introduced \cite{Jang}, which performs multiple local updates before synchronizing interface predictions. 
Also, this approach incorporates augmented Lagrange multipliers to dynamically balance Dirichlet boundary conditions and interface continuity constraints.
Compared to the previously mentioned methods, improved computational efficiency is achieved by reducing the need for computationally expensive global updates.
The newly introduced augmented Lagrange multipliers enhance convergence through the adaptive tuning of the penalty sMSE terms of the local loss functions.
Thus far, existing methods have been limited to physics-informed neural networks (PINNs), primarily aiming to enhance the convergence properties of increasingly complex models.

In this work, we propose more general, data-driven, non-overlapping domain decomposition approaches utilizing Lagrange multipliers to enforce interface constraints. 
Unlike the previously mentioned methods, our method also allows computing with data-driven NNs, utilizing parallel computations to allow the computation of large domains with local nonlinearities.
In addition to that, we add additional Lagrange multiplier terms to constrain Neumann boundary conditions for a smoother approximation, as previous works rely on a penalty-based MSE term for this.
One of the options is to remove the penalty terms altogether by using full Lagrange multipliers to get more precise interface approximations.
However, this comes at the cost of increased algorithmic complexity due to the required linearization of constraints at the interfaces.
Another option is to utilize augmented Lagrange multipliers, which are less computationally expensive.
These two methods are comprehensively tested on two example FEM simulations, comparing their performance and scalability for larger problems in design optimization.

 The paper is organized as follows: 
 In Section \ref{Sec:prob} the finite element equations are derived and NN surrogate models are introduced. 
 The DDMs with interface conditions are stated in Section \ref{Sec:method} and the two algorithms are presented. 
 Finally, in Section \ref{Sec:results} numerical results for a two-dimensional FEM simulation problem and a three-dimensional material optimization problem are shown and discussed.

\section{Problem statement}
\label{Sec:prob}
Consider a body occupying the spatial domain 
\(\gebiet \subset \mathbb{R}^n\), where 
\(x \in \gebiet\) denotes a material point within the body. The domain \(\gebiet\) has piecewise smooth Lipschitz boundaries, partitioned into Dirichlet boundary \(\partial \gebiet_D\) and Neumann boundary \(\partial \gebiet_N\) segments. Assuming quasi-static deformations and design parameters \(\zeta \in \mathbb{R}^m\), the following boundary value problem describes the mechanical behavior of the body under external forces:
\begin{align}
&\qquad &\nabla \cdot \sigma(\zeta, x) &= f(x), &\forall x& \in \gebiet,\label{Eq:PDE1}\\
    &\text{s.t.} &\quad u(x) &= u_D, &\forall x& \in \partial\gebiet_D,\label{Eq:PDE2}\\
    &\text{s.t.} \quad &\sigma(\zeta, x) \cdot n &= \tau_N, &\forall x& \in \partial\gebiet_N.
    \label{Eq:PDE3}
\end{align}
where $\sigma \in L_{2}(\gebiet, \textrm{Sym}(\mathbb{R}^{d}))$ is the stress tensor, $u \in \mathcal{U}$ with $\mathcal{U}:=\mathcal{H}^1_0(\gebiet)$ is the displacement field, $f \in \mathcal{U}^*$ (with $\mathcal{U}^*$ being the dual space of $\mathcal{U}$) describes the internal forces, $\tau_n$ denote Neumann boundary conditions on $\partial\gebiet_N$ and $u_D$ is the Dirichlet boundary condition on $\partial\gebiet_D$. 
The body is made of material undergoing deformation according to Hooke's constitutive law:
\begin{equation}
    \sigma = C:\epsilon,
    \label{Eq:Const}
\end{equation}
where \(C \in \mathcal{L}\big(\textrm{Sym}(\mathbb{R}^{d})\big)\) denotes the $4^{th}$ order symmetric, bonded, measurable and point-wise stable elasticity tensor, and strains $\epsilon \in L_{2}(\gebiet, \textrm{Sym}(\mathbb{R}^{d}))$ are defined as: 
\begin{equation}
    \epsilon = \frac{1}{2}(\nabla u +(\nabla u)^T).
    \label{Eq:linela}
\end{equation}
The strong formulation, as presented in Eqs.~(\ref{Eq:PDE1}–\ref{Eq:PDE3}), is rewritten in its weak form as follows:
\begin{equation}
    a(u,v):=\int_\gebiet C:\epsilon(u):\epsilon(v) dx = \int_\gebiet f(x) \cdot v(x) dx + \int_{\partial\gebiet_N} \tau_n(x) \cdot v(x) dA =: l(v),
    \label{Eq:weakf}
\end{equation}
in which $a(u,v)$ denotes the bi-linear form, and $l(v)$ is the corresponding linear functional. The problem has a unique solution according to the Lax-Milgram theorem \cite{Brezis2011}.

Subsequently, let the design parameters $\zeta$ be described probabilistically. In other words, let $\zeta$ be modeled as a random vector $\zeta(\omega)$ with a finite variance in a probability space $(S):=L_2(\varOmega,\mathcal{F},\mathbb{P})$ in which $\varOmega$ is the space of all events, $\mathcal{F}$ is the sigma algebra, and $\mathbb{P}$ is the probability measure.
Introducing $\zeta(\omega)$ in Eqs.~(\ref{Eq:PDE1}-\ref{Eq:PDE3}) rewrites the deterministic problem into its stochastic counterpart \cite{bojana}:
\begin{align}
&\qquad &\nabla \cdot \sigma(\zeta(\omega), x) &= f(x, \omega), &\forall x& \in \gebiet,\forall\omega \in \Omega,\label{Eq:PDEs1}\\
    &\text{s.t.} &\quad u(x,\omega) &= u_D, &\forall x& \in \partial\gebiet_D,\forall\omega \in \Omega,\label{Eq:PDEs2}\\
    &\text{s.t.} \quad &\sigma(\zeta(\omega), x) \cdot n &= \tau_N, &\forall x& \in \partial\gebiet_N,\forall\omega \in \Omega.
    \label{Eq:PDEs3}
\end{align}
in which $u \in \mathcal{U}\otimes(S)$. The constitutive law in Eq.~(\ref{Eq:Const}) becomes $\sigma(\omega) = C:\epsilon(\omega)$ with $\epsilon(\omega)$ defined as:
\begin{equation}
    \epsilon(\omega) = \frac{1}{2}(\nabla_s u(\omega) +(\nabla_s u(\omega))^T), \qquad \forall\omega \in \Omega.
    \label{Eq:linelastoch}
\end{equation}
Note, that one introduces the weak definition of the linear mapping $\nabla_s$ between the displacement $\epsilon \in \epsilon\otimes(S)$ and $u \in \mathcal{U}\otimes(S)$ such that:
\begin{equation}
\nabla_s: u_1(x)u_2(\omega) \rightarrow (\nabla_su_1(x))u_2(\omega),
\end{equation}
holds \cite{bojana}. Similarly to the deterministic problem, one can 
further derive a corresponding weak form of the equilibrium equation:
\begin{equation}
    \int_{\varOmega}\int_\gebiet C \epsilon(u(\omega))):\epsilon(v(\omega)) dx d\varOmega = \int_{\varOmega}\int_\gebiet f(x, \omega) \cdot v(x, \omega) dx d\varOmega + \int_{\varOmega}\int_{\partial\gebiet_N} \tau_n(x) \cdot v(x, \omega) dA d\varOmega, \label{Eq:swf}
\end{equation}
by integrating the terms over $\varOmega$.
As the solution of this initial value problem belongs to the tensor product space $\mathcal{U}\otimes(S)$, one requires the discretization of each of the spaces separately. The deterministic space $\mathcal{U}$ is discretized by using the finite element approach \cite{FEMbathe}:
\begin{equation}
    u_h:=span\{N_j(x)\}^{L_n}_{j=1},
\end{equation}
such that:
\begin{equation}
    u\vars \approx u_h\vars = \sum^{L_h}_{i=1} u_i(\omega)N_i(x),
    \label{Eq:usource}
\end{equation}
holds. Here, $N_i(x)$ denotes the shape function and $u_i(\omega)$ represents the unknown stochastic coefficient. Furthermore, the stochastic space is discretized by sampling with the Monte Carlo method \cite{Chan_2013}.
Hence, the problem given in Eq.~(\ref{Eq:swf}) is solved $n$ times in parallel for each realization of $\zeta(\omega_i)$, $i=1, \dots, n$.

As the Monte Carlo method is not sample-efficient, one may introduce a surrogate model to map the input parameters $(x,\zeta(\omega))$ to the QoI $\fvar(x, \zeta(\omega))$. In particular, this work focuses on a simple architecture, namely a feedforward neural network (NN). In other words, the discrete QoI $\fvar(x, \zeta(\omega))$ in Eq.~\ref{Eq:usource} is approximated by:
\beq
    \fvar\vars \approx \hat\fvar_{\ws}\vars =
    (\Fvar_L \circ \boldsymbol{a} \circ \Fvar_{L-1} \circ ... \circ \boldsymbol{a} \circ \Fvar_1)\vars, \qquad \ws := \left\{ \w^{(k)}, \be^{(k)} \right\}^L_{k=1},
    \label{Eq:compositionnn}
\eeq
in which $A_k:\real^{m_{k-1}} \rightarrow \real^{m_{k}}$ is an affine map, with $m_k \in \mathbb{N}$, defined as:
\beq
    x^{(k)} \mapsto \w^{(k)}x^{(k-1)} + b^{(k)}.
\eeq
Here, $\w^{(k)}$ and $b^{(k)}$ are the trainable network parameters collected in the vector $\ws$.
In Eq.~(\ref{Eq:compositionnn}), $\boldsymbol{a}$ represents the vector-valued function:
\beq
    \boldsymbol a (\boldsymbol h) := \left[a(h_1), a(h_2), ..., a(h_P)\right],
\eeq
acting element-wise on the hidden state $h_i$ of the NN (the output of the affine transformation before nonlinear activation function). $P$ and $L$ denote the number of nodes per layer and the number of layers, respectively. 

To determine the unknown parameters $\ws$ in the previously defined NN, we solve the following optimization problem:
\begin{equation*}
     \ws^* = \arg \min_\ws \loss(\ws).
\end{equation*}
Here, the objective function $\loss$ is typically defined in a mean-squared manner:
\begin{equation}
\mathcal{J}(\ws):=\mathbb{E}[ \| \fvar\vars-\hat\fvar_{\ws}\vars\|_2^2], \label{Eq:Jeq}
\end{equation}
in which, the expectation is approximated in a Monte Carlo simulation manner, by using a finite number of samples $n$. 
Finally, the approximation problem is solved by using an advanced gradient-based optimizer \cite{NoceWrig06}.

Obtaining a computationally feasible neural network surrogate model is challenging due to the complexity of the underlying problem. This complexity often results in a large number of trainable parameters, larger local approximation errors, and longer training times.
To overcome this issue, one may solve fixed number of local problems instead of the global one given in Eq.~(\ref{Eq:compositionnn}). 
The strategy proposed in this paper aims to reduce computational demands by partitioning the spatial domain. 
The resulting local problems are expected to be simpler to solve, require fewer parameters, and yield higher local accuracy.
\section{Domain decomposition}
\label{Sec:method}
Let the spatial domain $\gebiet$ be split into $M$ non-overlapping subdomains $\gebiet_i$, such that $\gebiet = \bigcup_{i=1}^{M} \gebiet_i$ and $\gebiet_i \cap \gebiet_j = \emptyset \ \text{for} \ i \ne j$. Then, the approximation, given in Eq.~(\ref{Eq:compositionnn}), can be substituted by 
a set of $M$ local NNs, each defined on its corresponding subdomain:
\begin{equation}
    \fvar_i\vars \approx \fwi\vars = (\Fvar_{L_i}^{(i)} \circ \boldsymbol{a}^{(i)} \circ \Fvar_{L_{i-1}}^{(i)} \circ ... \circ \boldsymbol{a}^{(i)} \circ \Fvar_1^{(i)})\vars, \qquad \forall x \in \gebiet_i, \qquad \forall \stochasticvar \in \varOmega, \label{Eq:splitnn}
\end{equation}
with
\beq
    \boldsymbol a^{(i)} (\boldsymbol y) = \left[a^{(i)}(y_1), a^{(i)}(y_2), ..., a^{(i)}(y_{P_i})\right].
\eeq
Here, $L_i$ and $P_i$ are the number of nodes per layer and the number of layers of the $i^{th}$ NN for each subdomain, respectively. In Eq.~(\ref{Eq:splitnn}), the function vector $\boldsymbol a^{(i)}$ acts element-wise on hidden states $y^{(i)}_1, \dots, y^{(i)}_{P_i}$ of each subdomain. Given the previously defined local NNs, the goal is to estimate the unknown parameters $\wi$, $i=1, \dots, M$
by minimizing the corresponding local objective:
\begin{eqnarray}
    \wi^*&=&\arg\min \mathcal{J}_i(\wi), \nonumber \\
    \quad  \loss_{\ixa}(\wi)&=&\mathbb{E} [\|\fwi\vars - \fvar\vars\|^2_2], 
    \quad \forall x \in \gebiet_i, 
    \quad \forall \omega \in \gebietf, 
    \quad i,...,M, \label{Eq:unconst_eq_loc}
\end{eqnarray}
given the local dataset :
\[
\mathcal{D}_i := \{ ((\point_i, \zeta(\omega)), \fvar(\point_i, \zeta(\omega)) \}_{i=1}^{M}, \quad \text{with } \point_i \in \gebiet_i, \quad \forall i=1,\dots,M, \quad \stochasticvar \in \gebietf,
\]
where \(m_i\) denotes the number of spatial samples in the \(i\)-th subdomain \(\gebiet_i\).

The previously discussed local approximations suffer from discontinuities at the interfaces between subdomains. To overcome this issue, $C^0$ or $C^1$ continuity constraints must be imposed on the interfaces between the subdomains to ensure a continuous approximation. In this paper we extend the approach to account for $C^1$ continuity. In other words, both the function and its normal gradients must be continuous across the interface of adjacent subdomains:
\begin{align}
    \fwi(x, \zeta(\omega)) &= \fm(x, \zeta(\omega)), &\quad \forall x \in \bound, &\quad \stochasticvar \in \gebietf,
    \label{Eq:valcons} \\
    \partial_\normal \fwi(x, \zeta(\omega))|_\normal &=  \partial_\normal \fm(x, \zeta(\omega))|_\normal, &\quad \forall x \in \bound, &\quad \stochasticvar \in \gebietf,
    \label{Eq:gradcons}
\end{align}
where $\normal$ is normal to the interface, $\fm$ is the interface solution defined on the interface $\bound = \gebiet_{\ixa} \cap \gebiet_{\ixb}$, $\ixa \neq \ixb$. Note, that the solution on the interface $\fm$ can be also approximated by its own local NN. 

By introducing the interface constraints given in Eqs.~(\ref{Eq:valcons}-\ref{Eq:gradcons}) to the local optimization problems in Eq.~(\ref{Eq:unconst_eq_loc}), one may rewrite the global approximation problem in Eq.~(\ref{Eq:Jeq}) by the local ones:
\begin{align}
    &\qquad &\wi = \arg\min_{\wi} &\loss_{\ixa}(\wi),\label{Eq:unconst_eq_loc1} &i=1, \dots, M,\\
    &\text{s.t.} &\quad \fwi(x, \zeta(\omega)) &= \fm(x, \zeta(\omega)), &\forall x \in \bound , & \quad \forall \omega \in \gebietf, \label{Eq:unconst_eq_loc2} \\
    &\text{s.t.} \quad &\partial_\normal \fwi(x, \zeta(\omega))\big|_\normal &= \partial_\normal \fm(x, \zeta(\omega))\big|_\normal, &\forall x \in \bound , & \quad \forall \omega \in \gebietf. \label{Eq:unconst_eq_loc3}
\end{align}

At the same time, one has to solve the interface condition for two adjacent domains $\gebiet_\ixa$ and $\gebiet_\ixb$ with solutions $\hat\fvar_{\ws_{\ixa}}$ and $\hat\fvar_{\ws_{\ixb}}$. In other words, one has to solve the optimization problem at the interface:
\begin{equation}
    \ws_{ik}^* = \arg\min \mathcal{J}_{\ixa\ixb}(\ws_{\ixa\ixb})
\end{equation}
in which
\begin{equation}
    \mathcal{J}_{\ixa\ixb}(\ws_{\ixa\ixb}) = \sum_{\{\ixa,\ixb\}} \mathbb{E} \Big[
        \|\fwj(x, \zeta(\omega)) - \fm(x, \zeta(\omega))\|^2_2 \Big] 
        + \sum_{\{\ixa,\ixb\}} \mathbb{E} \Big[
        \|\partial_\normal \fwj(x, \zeta(\omega))|_\normal - \partial_\normal \fm(x, \zeta(\omega))|_\normal\|^2_2 \Big], \quad \forall \point \in \bound.
        \label{Eq:inbetween}
\end{equation}
For simplicity, the following discussion focuses solely on interfaces between pairs of adjacent domains, which can then easily be extended to multiple interfaces.

\subsection{Lagrange multiplier algorithm}
\label{Sec:method1}
To reformulate Eqs. (\ref{Eq:unconst_eq_loc1}-\ref{Eq:unconst_eq_loc3}) as an unconstrained optimization problem, the method of Lagrange multipliers is introduced:
\begin{equation}
    \Lagr_\ixa (\wi ,\li) = 
    \loss_{\ixa}(\wi)
    +\li^T \Cwi,
    \label{Eq:main_eq}
\end{equation}
with
\begin{equation}
    \Cwi
    :=
    \begin{bmatrix}
        \fwi(x, \zeta(\omega)) - \fm(x, \zeta(\omega)) = 0\\
        \partial_\normal \fwi(x, \zeta(\omega))|_\normal -  \partial_\normal \fm(x, \zeta(\omega))|_\normal = 0\\
    \end{bmatrix}, \qquad \forall\point \in \bound, \qquad \forall\omega \in \gebietf,
    \label{Eq:const_mat}
\end{equation}
and \(\boldsymbol{\lambda}_i\) being the multipliers. The formulation leads naturally to a non-linear unconstrained optimization problem, since each local subdomain is represented by a NN and therefore introduces strong, highly non-convex dependencies on the trainable parameters \(\wi\) and Lagrange multipliers \(\li\).
Nonlinear problems are commonly solved sequentially by local approximations, e.g., as quadratic problem with linear constraints, i.e., Sequential Quadratic Programming (SQP) \cite{NoceWrig06}.
Although SQP can be powerful in solving NNs, it is not available with constraints in common NN packages (PyTorch \cite{pytorch}, TensorFlow \cite{tensorflow}). 
In SQP one would define the dual functions: 
\begin{equation}
    g_i(\li)=\inf_\wi \Lagr_i(\wi,\li), i=1, \dots, M, 
\end{equation}
which give a lower bound on the primal objective for any \(\li\), and hence one formulates the dual problems:
\begin{equation}
\max_\li g_i(\li),
\end{equation}
solved by gradient ascent.
Ideally, a SQP method with constraints is used, as it enforces the constraints intrinsically during each update, which is generally more efficient than applying constraint checks or corrections a posteriori.
However, due to the lack of Python packages, a first method checking the constraints after updating is taken to solve such nonlinear problems.
Instead, dual ascent~\cite{Boyd2010} is used. Dual ascent iterates the dual and primal updates:
\begin{align}
\wi^{\iter+1} &:= \arg\min_{\wi} \, \Lagr_i(\wi, \li^\iter) \label{eq:update_wi}, \\
\li^{\iter+1} &:= \li^\iter + \alpha \Constr(\wi^{\iter+1}) \label{eq:update_li},
\end{align}
$\iter, \dots, D_m$ times, in which $\alpha$ is the learning rate. 
The remaining problem is to define a convergent primal update with the local MSE term and the interface constraints as shown in Eq.~(\ref{Eq:main_eq}).
Directly solving the problem remains challenging as NNs are highly nonlinear and almost always non-convex.
The dual ascent algorithm requires strong duality, i.e., satisfaction of the Slater condition \cite{Slater}.
To address this, the constraints in Eq.~(\ref{Eq:const_mat}) are linearized iteratively around reference points $\hat\ws_i$ using Taylor expansions:
\begin{equation}
    \bar{\fvar}_\wi(x, \zeta(\omega)) = \hat\fvar_{\hat\ws_i}(x, \zeta(\omega)) + \partial_\wi \fwi(x, \zeta(\omega))|_{\hat\ws_i} (\wi - \hat\ws_i), 
    \qquad \forall x \in \bound, \qquad \forall \omega \in \gebietf,
    \label{Eq:lin1}
\end{equation}
and similarly for its normal derivative,
\begin{equation}
    \partial_\normal \bar{\fvar}_\wi|_{\normal}(x, \zeta(\omega)) = \partial_\normal \hat\fvar_{\hat\ws_i}(x, \zeta(\omega))|_{\normal} + 
    \partial_{\normal}\partial_{\wi} \fwi(x, \zeta(\omega))|_{\hat\ws_i, \normal} (\wi - \hat\ws_i), 
    \qquad \forall x \in \bound, \qquad \forall \omega \in \gebietf.
    \label{Eq:lin2}
\end{equation}
Substituting Eqs.~(\ref{Eq:lin1}-\ref{Eq:lin2}) into Eq.~(\ref{Eq:main_eq}) yields
\begin{equation}
    \bar{\Lagr}_i (\wi) = \loss_{\ixa}(\wi) + (\li^\iter)^T \Cwj,
    \label{Eq:ffinal}
\end{equation}
with
\begin{equation}
    \Cwj =
    \begin{bmatrix}
        \bar{\fvar}_\wi(x, \zeta(\omega)) - \fm(x, \zeta(\omega))\\
        \partial_\normal \bar{\fvar}_\wi(x, \zeta(\omega))|_{\normal} - \partial_\normal \fm(x, \zeta(\omega))|_\normal
    \end{bmatrix}, \qquad \forall \point \in \bound.
    \label{Eq:cfinal}
\end{equation}
The problem that arises is a non-convex objective function with a linearized constraint, leading to local minima, as usually encountered in NN literature \cite{GoriTesi1992}.
The linearization is assumed to be accurate enough, given that the Lagrange update step is sufficiently small and the parameter space around the previous Lagrange multiplier $\li^\iter$ also exhibits local linearity.
Under these conditions, gradient-based methods can converge in $\wi$ by repeatedly minimizing Eq.~(\ref{Eq:lin1}) at updated linearization points.  
If the updated parameters fall outside the feasible region, the last feasible parameters (where $\Cwi = \Cwj$) are used as the new linearization point.  
This sequentially linearized constraint problem is solved using a gradient-based optimization. More precisely, the Limited-memory Broyden–Fletcher–Goldfarb–Shanno (LBFGS) algorithm~\cite{lbfgs} is used as optimizer.  
LBFGS is suitable in this context since the decomposition leads to inherently small local problems with relatively few network parameters, enabling efficient Hessian approximations. 
Furthermore, compared to first-order methods, L-BFGS provides a more accurate update direction, which helps maintain the constraint feasibility condition $\Cwi = \Cwj$.  
In summary, the quasi-Newton method L-BFGS is used to sequentially update the network parameters within the linearized constraint problems.
Once the dual ascent algorithm of all local domains is converged, the interface NN is trained by LBFGS to update the interface predictions.
These two steps are repeated sequentially until the method is fully converged.

Reformulating the governing equations as an unconstrained Lagrange problem leads to the underlying algorithmic structure of Alg.~\ref{Al:DDM}. 
Initially, the spatial domain is decomposed into a set of subdomains:
\[
\mathcal{D} = \{\gebiet_1,\dots,\gebiet_M\},
\]
which are connected through interfaces:
\[
\mathcal{I} = \bigcup_{\ixa<\ixb} \partial \bound.
\]
In each subdomain $\gebiet_i$, a local NN with parameters $\wi$ is defined to approximate the solution. In addition, interface NNs with parameters $\wm$ are introduced to represent the interfaces between neighboring subdomains. 
All local and interface network parameters are initialized using a He~Normal distribution, while the Lagrange multipliers $\li$ associated with the interface constraints are initialized as zero. The algorithm proceeds iteratively for $k = 0,1,\dots,K_m$.
At each iteration, the interface networks are first predicted at a discrete set of interface points $\point \in \partial \gebiet_{\ixa\ixb}$. These evaluations yield the interface predictions $\fm^{(k)}(\point)$ and their normal derivatives $\partial_\normal \fm^{(k)}(\point)$, which serve as reference predictions to enforce interface continuity. 
These reference predictions are substituted into Eq.~(\ref{Eq:const_mat}) to make the constraints solely dependent on local network parameters $\wi$.
Each local constraint optimization problem represented by a NN is minimized with respect to the local loss $\Lagr_{\ixa}(\wi)$ in Eq.~(\ref{Eq:main_eq}) subject to the interface constraints. . 
These local problems are addressed using a dual ascent strategy as presented later in Alg.~\ref{Al:DA}, in which the primal variables $\wi$ and the corresponding Lagrange multipliers $\li$ are updated iteratively.
After updating all subdomain problems, the interface networks are re-trained by fixing the updated local network parameters $\wi^{(k+1)}$ and minimizing an interface loss function $\loss_{\ixa\ixb}$ in Eq.~(\ref{Eq:inbetween}) using an L-BFGS optimizer. This step ensures consistency between the interface networks and the local solutions.
Convergence is assessed using three criteria, primal stationarity, dual stationarity, and fulfillment of the interface constraints. Primal stationarity requires the expected norm of the gradient of the local Lagrangian with respect to $\wi$ to fall below a prescribed tolerance. Dual stationarity ensures that changes in the constraint residuals between successive iterations are sufficiently small to be below a tolerance. Constraint fulfillment requires the expected magnitude of the interface constraint residuals to be below a tolerance. If all criteria are satisfied, the algorithm terminates and returns the converged local network parameters $\wi^*$.

\begin{algorithm}[ht!]
\caption{Substructuring domain decomposition of NNs}
\begin{algorithmic}[1]
\Require Subdomains: $\mathcal{D}=\{\gebiet_1,\dots,\gebiet_M\}$, Interfaces: $\mathcal{I}=\bigcup_{\ixa<\ixb}\bound$
\Statex\quad\quad\quad Parameters: $\alpha, \ \varepsilon_{\ixa\ixb}, \ \varepsilon_{pr}, \ \varepsilon_\lambda , \ \varepsilon_l,\ K_m,\ D_m,\ P_m$ 
\Statex\quad\quad\quad Functions: $Q_i(\cdot,\cdot), \loss_\ixa(\cdot), \loss_{\ixa\ixb}(\cdot, \cdot)$
\Statex
\State \textbf{Initialize:} $\wi^{(0)},\wm^{(0)}\sim\mathrm{HeNormal},\quad \li^{(0)}=\boldsymbol{0}\quad \forall i=1,\dots,M$

\For{$k=0,1,2,\dots,K_m$}

  \ForAll{interface $\bound\in\mathcal{I}$}
    \State Evaluate interface NN at discrete points $\point \in \bound$:
    \[
    \fm^{(k)}(\point)=\fp_{\ixa\ixb}^{(k)}(\point;\wm^{(k)}), \quad 
    \partial_\normal \fm^{(k)}(\point)|_{\normal}=\partial_\normal \fp_{\ixa\ixb}^{(k)}(\point;\wm^{(k)})|_{\normal},
    \quad \forall \point \in \bound
    \]
  \EndFor
  
  \ForAll{subdomains $\gebiet_i\in\mathcal{D}$}
    \State Define interface constraint function for fixed $\wm^{(k)}$:
    \[
    \Constr(\cdot,\wm^{(k)}):=
    \begin{bmatrix}
    \fwi(\point)-\fm^{(k)}(\point), \\
    \partial_\normal \fwi(\point)\big|_{\normal}
    - \partial_\normal \fm^{(k)}(\point)\big|_{\normal}
    \end{bmatrix}
    \]
    \State as in Eq.~\ref{Eq:const_mat} and the sub-domain MSE loss $\loss_\ixa(\cdot)$ according to Eq.~(\ref{Eq:unconst_eq_loc}).
    \State Solve local constrained problem with Alg.~\ref{Al:DA} for each sub-domain:
    \[ (\wi^{(k+1)},\li^{(k+1)}) = \mathrm{DUAL\ ASCENT}(\wi^{(k)},\li^{(k)},D_m,P_m,\varepsilon_{pr},\varepsilon_\lambda,\varepsilon_l,\Constr(\cdot,\wm^{(k)}),\loss_\ixa(\cdot), \alpha) \]
  \EndFor
    
  \ForAll{interface $\bound\in\mathcal{I}$}
    \State Fit interface NN using Eq.~(\ref{Eq:inbetween}) to find the interface parameters $\wm$ by fixing $\wi$ to $\wi^{(k+1)}$:
    \[ \wm^{(k+1)} = \mathrm{LBFGS}(\loss_{\ixa\ixb}(\cdot,\wi^{(k+1)}),\ \varepsilon_{\ixa\ixb}) \]
  \EndFor

  \Statex

  \State \textbf{Convergence check:}
  \State \qquad Primal stationarity:
  \[
  \mathbb{E}\left[\left|\nabla_{\theta_i}\mathcal{L}_i(\wi^{(k+1)},\li^{(k+1)})\right|\right] \le \varepsilon_{pr},\ \forall i,
  \]
  \State \qquad Dual stationarity:
  \[
  \mathbb{E}\left[\left|\Constr(\wi^{(k+1)}, \wm^{(k+1)})-\Constr(\wi^{(k)}, \wm^{(k+1)})\right|\right]
  \le \varepsilon_\lambda,\ \forall i, 
  \]
  \State \qquad Constraint fulfillment (interfaces):
  \[
\mathbb{E}\left[\left|\Constr(\wi^{(k+1)},\wm^{(k+1)})\right|\right] \le \varepsilon_{\ixa\ixb},\ \forall \ixa,\ixb, 
  \]
  \If{Convergence check satisfied}
        \State Go to return
  \EndIf
\EndFor

\State \textbf{Return:} $\wi^*=\wi^{(k+1)}$
\end{algorithmic}
\label{Al:DDM}
\end{algorithm}

Algorithm~\ref{Al:DA} follows a primal--dual optimization strategy based on dual ascent.
Given the current iterate $(\ws^{(d)},\ls^{(d)})$, the primal variables are updated first by minimizing the Lagrangian with respect to $\ws$ while holding the dual variables $\ls$ fixed.
The objective function $\Lagr_i$ is then minimized solely with respect to the primal variables $\wi$. 
If the multipliers are zero, the update corresponds to solving the unconstrained problems defined by $\loss_i$ in Eq.~(\ref{Eq:Jeq}). 
Otherwise, the primal update is carried out for the constrained problem, which includes the constraint term $\Constr$ in addition to $\loss_i$ as in Eq.~(\ref{Eq:main_eq}).
Primal update functions are defined in detail later in Algs.~\ref{Al:DA_nl} and \ref{Al:DA_al}.
Once the primal variables have been updated, the dual update adjusts the Lagrange multipliers according to Eq.~(\ref{eq:update_li}).
These two steps are repeated until both parameter sets have converged, that is, until the constraint term $\Constr$ and the gradient of the objective with respect to the network parameters, $\nabla_{\wi}\mathcal{L}(\wi,\li)$, vanish within a prescribed tolerance.

\begin{algorithm}[t!]
\caption{Dual Ascent of NNs}
\begin{algorithmic}[1]
\Function{DUAL ASCENT}{$\boldsymbol{\xi}, \boldsymbol{\zeta}, D_m, P_m, \varepsilon_\lambda, \varepsilon_{pr}, \varepsilon_l, \Q(\cdot), \loss(\cdot), \alpha$}

    \State Initialize: $\ws^{(0)} = \boldsymbol{\xi}$, \; $\ls^{(0)} = \boldsymbol{\zeta}$
    \For{$d = 0,1,2,\dots,D_m$}
        \State \textbf{Primal update:}
        \If{$\ls = \boldsymbol{0}$}
        \State Fit local NN according to MSE loss $\loss$ from Eq.~(\ref{Eq:unconst_eq_loc}):
        \[
        \ws^{(d+1)} = \mathrm{LBFGS}(\loss(\cdot), \ \varepsilon_{pr}), 
        \]
        \State and set $D_m = 0$.
        
        \Else
        \State Fix $\ls$ to $\ls^{(d)}$ and solve for $\ws$: 
        \[
        \ws^{(d+1)} = \mathrm{PRIMAL\ UPDATE}(\ws^{(d)}, \ls^{(d)}, P_m, \varepsilon_{pr}, \varepsilon_l,\Q(\cdot), \loss(\cdot), \alpha).
        \]
        
        \EndIf
        \State \textbf{Dual update:}
        \State Update lagrange multipliers according to Eq.~(\ref{eq:update_li}):
        \[
            \ls^{(d+1)} = \ls^{(d)} + \alpha \, \Q(\ws^{(d+1)}).
        \]
        \State \textbf{Convergence check:}
        \[
            \mathbb{E}\left[\left| \Q(\ws^{(d+1)}) \right|\right] < \varepsilon_\lambda,
        \]
        \[
            \mathbb{E}\left[\left| \nabla_{\ws} \mathcal{L}(\ws^{(d+1)}, \ls^{(d+1)}) \right|\right] < \varepsilon_{pr},
        \]
        \If{Convergence check satisfied}
        \State Go to return
        \EndIf
    \EndFor
    \State \Return $\ws^* = \ws^{(d+1)}$, \; $\ls^* = \ls^{(d+1)}$
\EndFunction
\end{algorithmic}
\label{Al:DA}
\end{algorithm}

An overview of the primal update procedure using LBFGS with linearized constraints is given in Alg.~\ref{Al:DA_nl}, which is referred to as Lagrange multiplier algorithm (LMA). 
The algorithm takes the local NN parameters and the corresponding Lagrange multipliers as inputs, where the multipliers remain fixed throughout the primal update. 
Each iteration begins by linearizing the constraints according to Eq.~(\ref{Eq:cfinal}), which yields the linearized loss in Eq.~(\ref{Eq:ffinal}). 
This linearized loss is then minimized using LBFGS with gradient:
\begin{equation}
    \nabla_{\ws} \bar{\Lagr}(\ws,\ls)
    = \nabla \loss(\ws) + J_Q(\ws^{(p)})^\top \ls,
\end{equation}
where $J_Q(\ws^{(p)})$ denotes the Jacobian of the constraint evaluated at iteration $p$. 
The expression shows that the linearization introduces a constant gradient shift induced by the dual variables, without modifying the Hessian approximation used by LBFGS. The linearization and optimization steps are repeated while the linear constraint approximation remains valid and until the network parameters have converged.

\begin{algorithm}[ht]
\caption{Primal update of dual ascent with LBFGS and Linearized Constraints}
\begin{algorithmic}[1]
\Function{PRIMAL UPDATE}{$\boldsymbol\xi, \boldsymbol\zeta, P_m, \varepsilon_{pr}, \varepsilon_l, \Q(\cdot), \loss(\cdot), \rho$}
    \State Initialize: $\ws^{(0)} = \boldsymbol\xi$, \; $\ls = \boldsymbol\zeta$
    \For{$p = 0,1,2,\dots,P_{m}$}
        \State Linearize constraints w.r.t. $\ws$ as stated in Eqs.~(\ref{Eq:lin1},\ref{Eq:lin2},\ref{Eq:cfinal}) for a given $\ws^{(p)}$:
        \[
            \Q(\ws) \approx \Q(\hat\ws^{(p)}) 
            + J_{\Q}(\hat\ws^{(p)})(\ws - \hat\ws^{(p)})
            =: \bar{\Q}(\ws).
        \]
        \State Update network parameters $\ws$ according to the loss defined in Eq.~(\ref{Eq:ffinal}):
        \[
            \mathcal{L}_i(\ws, \ls) 
            \approx \loss(\ws) + \ls^\top \bar{\Q}(\ws)
            =: \bar{\mathcal{L}}(\ws, \ls),
        \]
        
        \State by finding updated value of $\ws$ with a fixed $\ls$ using:
        \[
        \ws^{(p+1)} = \mathrm{LBFGS}(\bar{\mathcal{L}}(\cdot, \ls), \ \varepsilon_{pr}), 
        \]
        \State with loss $\bar{\mathcal{L}}$ gradient:
        \[
            \nabla_{\ws} \bar{\mathcal{L}}(\ws, \ls)
            = \nabla \mathcal{J}(\ws) 
            + J_Q(\ws^{(p)})^\top \ls.
        \]
        \State \textbf{Convergence check :}
        \State \qquad Primal stationarity:
        \[
            \mathbb{E}\left[\left| \nabla_{\ws} \mathcal{L}(\ws^{(p+1)}, \ls) \right|\right] < \varepsilon_{pr},
        \]
        \State \qquad Linearization feasibility:
        \[
            \mathbb{E}\left[\left|\bar{\Q}(\theta^{(p+1)})-\Q(\ws^{(p+1)})\right|\right] < \varepsilon_l,
        \]
        \If{Convergence check satisfied}
        \State Go to return
        \EndIf
    \EndFor
    \State \Return $\ws^* = \ws^{(p+1)}$
\EndFunction
\end{algorithmic}
\label{Al:DA_nl}
\end{algorithm}

\newpage
\subsection{Augmented Lagrange multiplier algorithm}
\label{Sec:method2}
Since the linearization of NN predictions is tedious and computationally expensive, one can also reformulate the objective function in Eqs. (\ref{Eq:unconst_eq_loc1}-\ref{Eq:unconst_eq_loc3}) by using the augmented Lagrange multipliers $\li$ form \cite{Boyd2010}:
\begin{equation}
    \Lagr_\ixa (\wi ,\li) = 
    \loss_{\ixa}(\wi)
    +\li^T \Cwi+\pen\|\Cwi\|^2,
    \label{Eq:main_eq2}
\end{equation}
with
\begin{equation}
    \Cwi
    :=
    \begin{bmatrix}
        \fwi(\point) - \fm(\point) = 0\\
        \partial_\point \fwi(\point)|_\point -  \partial_\point \fm(\point)|_\point = 0\\
    \end{bmatrix}, \qquad \forall\point \in \bound.
    \label{Eq:const_mat2}
\end{equation}

Here, an additional regularization term ($\|\Cwi\|^2$), in this case of a sum squared type, is used to penalize growth of the constraints in Eq. (\ref{Eq:main_eq2}).
The network parameters $\wi$ and Lagrange multipliers $\li$ are updated in a similar manner to Eqs. (\ref{eq:update_wi} - \ref{eq:update_li}), by gradient-based methods in the previous section.
In case of $\li$, the iterative updates:
\beq
\li = \li + \pen \Cwi,
\label{Eq:lambda_update}
\eeq
are dependent on the penalty parameter $\pen$ as step size (learning rate) and utilize steepest gradient ascent to ensure that the optimization problem is bounded properly.
Given these adjusted equations, one can write the iterative update:
\begin{align}
\wi^{\iter+1} &:= \arg\min_{\wi} \, \Lagr_i(\wi, \li^\iter), \label{eq:update_wi_al} \\
\li^{\iter+1} &:= \li^\iter + \pen \Cwiii, \label{eq:update_li_al}
\end{align}
which can directly be solved by gradient-based methods without further adjustments.

The resulting augmented Lagrange multiplier algorithm (ALMA), which utilizes LBFGS for the primal update is presented in Alg.~\ref{Al:DA_al}. 
Similar to LMA, the ALMA primal update function also gets as inputs the to be updated network parameters and fixed Lagrange multipliers.
Then, LBFGS updates are repeated on the loss of Eq.~(\ref{Eq:main_eq2}) until the network parameters have converged.

\begin{algorithm}[ht]
\caption{Primal update of dual ascent with LBFGS and augmented Lagrange multipliers}
\begin{algorithmic}[1]
\Function{PRIMAL UPDATE}{$\boldsymbol\xi, \boldsymbol\zeta, P_m, \varepsilon_{pr}, \varepsilon_l, \Q(\cdot), \loss(\cdot), \rho$}
    \State Initialize: $\ws^{(0)} = \boldsymbol\xi$, \; $\lambda = \boldsymbol\zeta$
    \For{$p = 0,1,2,\dots,P_{m}$}
        \State Update network parameters $\ws$ according to the loss defined in Eq.~(\ref{Eq:main_eq2})
        \[
            \mathcal{L}(\ws, \ls) 
            = \loss(\ws) + \lambda^\top \Q(\ws)
            + \rho \Q(\ws)^T \Q(\ws)
        \]
        \State by finding updated value of $\ws$ with a fixed $\ls$ using:
        \[
        \ws^{(p+1)} = \mathrm{LBFGS}(\mathcal{L}(\cdot, \ls), \ \varepsilon_{pr}), 
        \]
        \State \textbf{Convergence check :}
        \State \qquad Primal stationarity:
        \[
            \mathbb{E}\left[\left| \nabla_{\ws} \mathcal{L}(\ws^{p+1}, \ls) \right|\right] < \varepsilon_{pr}
        \]
        \If{Convergence check satisfied}
        \State Go to return
        \EndIf
    \EndFor
    \State \Return $\ws^* = \ws^{(p+1)}$
\EndFunction
\end{algorithmic}
\label{Al:DA_al}
\end{algorithm}

\newpage
\section{Results}
\label{Sec:results}
In this section, several numerical experiments are conducted to evaluate the proposed methods. 
The first example, in Section \ref{Sec:2D}, considers a two-dimensional cylinder under compression, which is used to compare the accuracy and computational efficiency of the two NN DDM algorithms by splitting into multiple domains. 
In addition, \al is analyzed with respect to its sensitivity to problem properties, specifically the number of sampling points per domain and the presence of gaps between local domains. 
As a second example, in Section \ref{Sec:3D}, a three-dimensional cylinder compression test with uncertainty propagation of the linear-elastic material parameters, i.e., the bulk and shear moduli is examined. 

\subsection{Compression test of 2D cylinder}
\label{Sec:2D}
\begin{figure}[ht!]
        \centering
        \begin{subfigure}[b]{0.2\textwidth}
           \centering
            \includegraphics[width=0.6\textwidth]{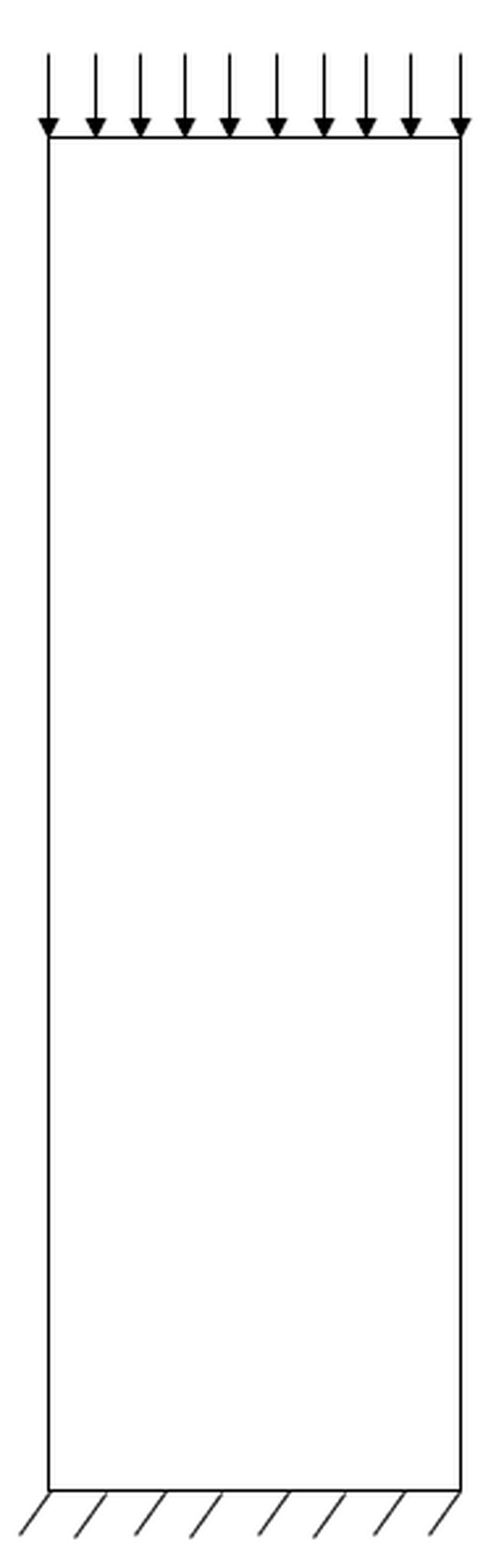}
            \caption{}     
            \label{Fig:rectangle}
        \end{subfigure}
        \raisebox{-2mm}{
        \begin{subfigure}[b]{0.2\textwidth}  
            \centering
            \includegraphics[width=1.3\textwidth]{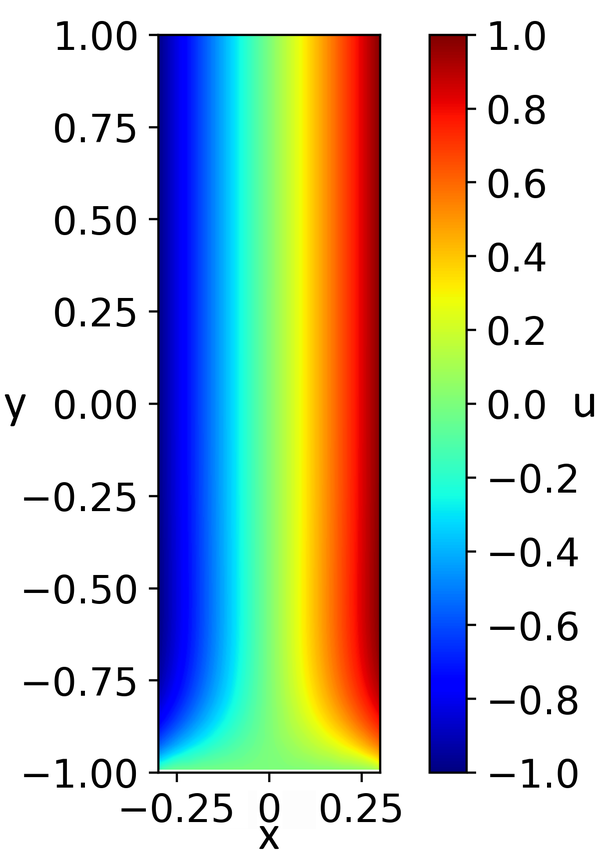}
            \caption{}     
            \label{Fig:int_rect}
        \end{subfigure}
        \hspace{10mm}
        \begin{subfigure}[b]{0.2\textwidth}  
            \centering
            \includegraphics[width=0.88\textwidth]{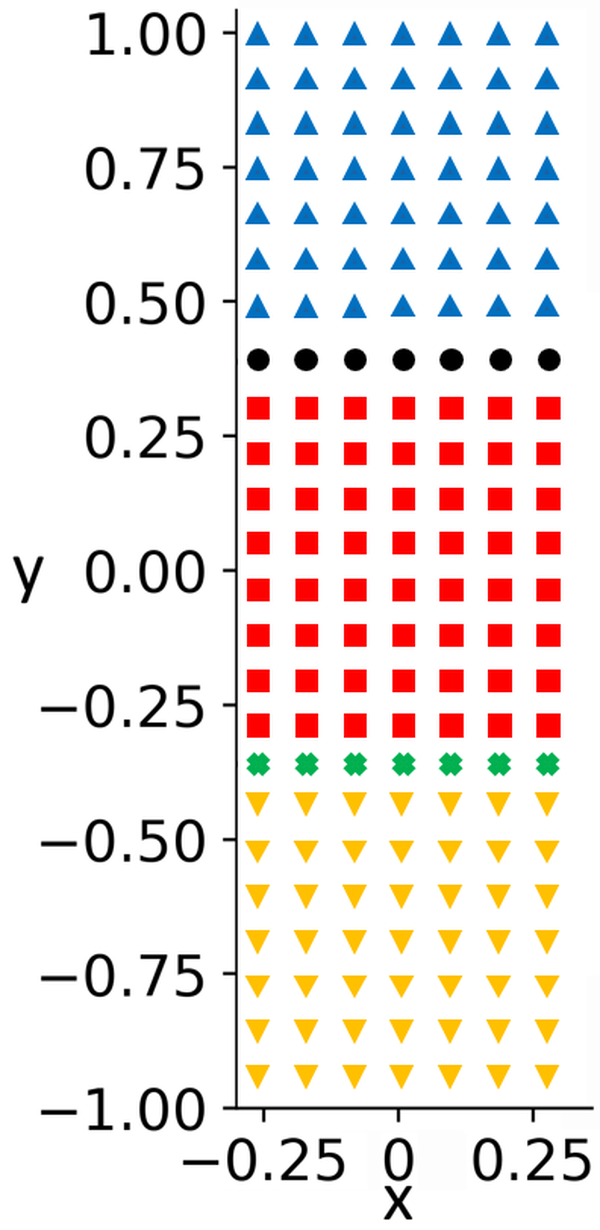}
            \caption{}     
            \label{Fig:points}
        \end{subfigure}
        }
        \caption{a) Schematic of the 2D specimen clamped at the bottom and compressed by a predefined displacement at the top. b) Interpolation plot of the normalized horizontal displacement field of the specimen. c) Schematic view of possible split sampling domains (3) from top to bottom, in blue, red and yellow, with in green and black possible interface points.} 
        \label{Fig:2D_problem}
    \end{figure}

\begin{table}[b!]
    \centering
    \caption{Specimen properties for the finite element simulation.}
    \begin{tabular}{llllllllll}
    \toprule
        \multicolumn{3}{c}{Dimensions [mm$^2$]} & \multicolumn{1}{c}{Young's} & \multicolumn{1}{c}{Poisson} & \multicolumn{1}{c}{Element} & \multicolumn{2}{c}{Number of}\\ \cmidrule(r){1-5} \cmidrule(r){9-10}
         \multicolumn{1}{c}{width} & x & \multicolumn{1}{c}{height} & \multicolumn{1}{c}{mod [MPa]} & \multicolumn{1}{c}{ratio} & \multicolumn{1}{c}{type} & \multicolumn{1}{c}{elements} &\multicolumn{1}{c}{nodes}\\ \midrule
        21 & x & 70 & 21000 & 0.3 & planar shell & 1800 & 234 x 71\\
    \bottomrule
    \end{tabular}
    
    \label{Tab:rectangle}
\end{table}

The problem setup shown in Fig.~\ref{Fig:rectangle} considers a two-dimensional specimen made of a linear elastic material. 
The specimen is clamped along the bottom edge and subjected to a prescribed vertical displacement of 5 mm along the top edge. 
The specimen's geometric and material properties are summarized in Tab.~\ref{Tab:rectangle}.
After FEM discretization by using 1800 two-dimensional plane strain elements, the horizontal displacement field is extracted as the QoI, while the corresponding spatial coordinates serve as the two-dimensional input feature data. 
The data is sampled from the nodal points of the FEM discretization, similarly to the illustration in Fig.~\ref{Fig:points}, which is schematically split into three domains based on the given grid with interface points in between the domains.
Both inputs and outputs are normalized, as illustrated in Fig.~\ref{Fig:int_rect}, to enable consistent testing of the proposed DDMs. Based on this normalized dataset, several experiments are conducted to evaluate the performance of both \nl and \al, which are discussed in detail in the following sections. For all computations single Linux cluster nodes made of $32$ Intel(R) Xeon(R) Silver $4216$ $2.10$ GHz CPU processors with a total of $140$ GB memory are used.

\begin{figure}[b!]
    \centering

    \begin{subfigure}[t]{0.475\textwidth}
        \centering
        \includegraphics[width=\linewidth]{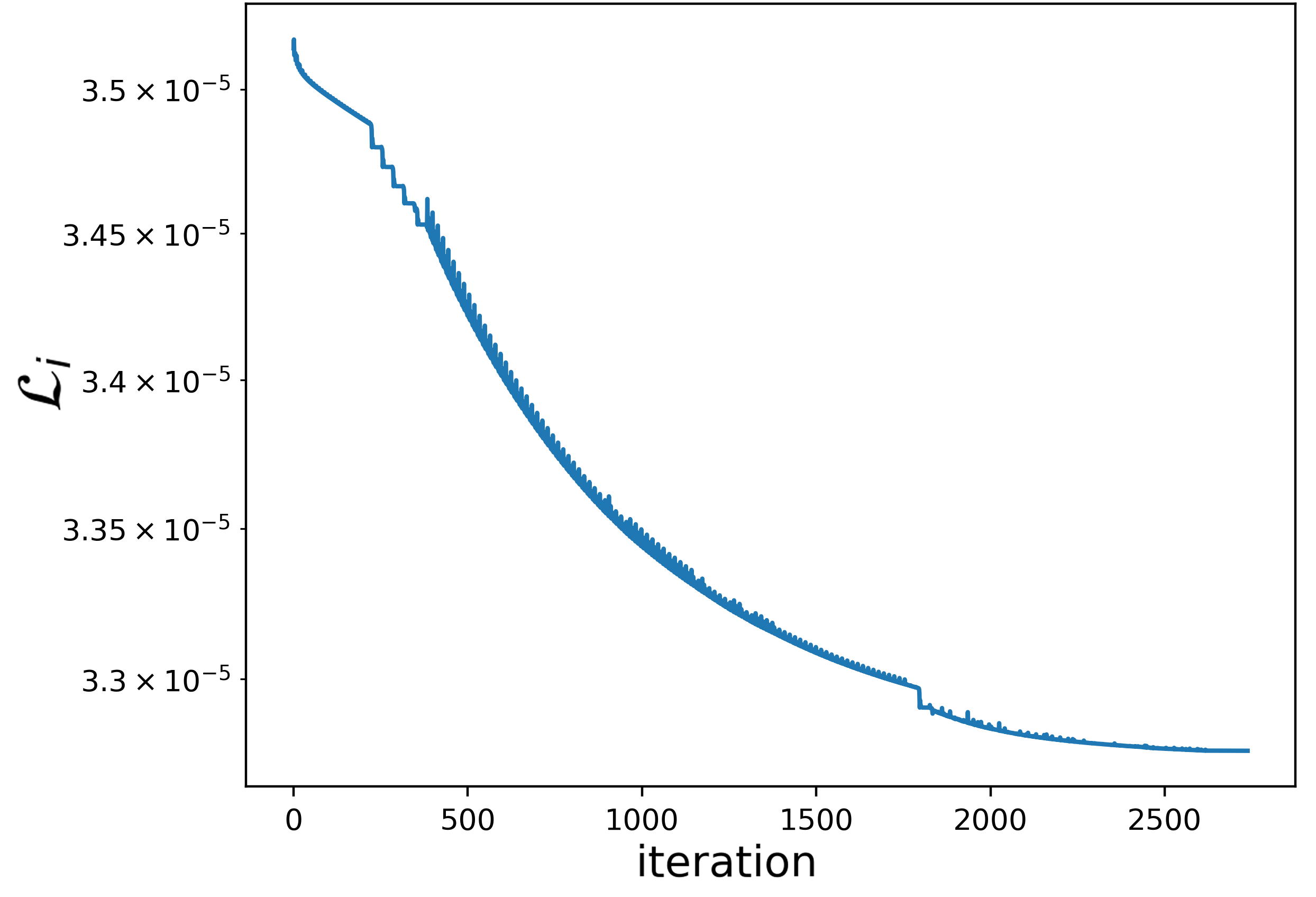}
        \caption{Convergence primal update \nl iteration 2.}
        \label{Fig:conv_inner_nl}
    \end{subfigure}
    \hfill
    \begin{subfigure}[t]{0.475\textwidth}
        \centering
        \includegraphics[width=\linewidth]{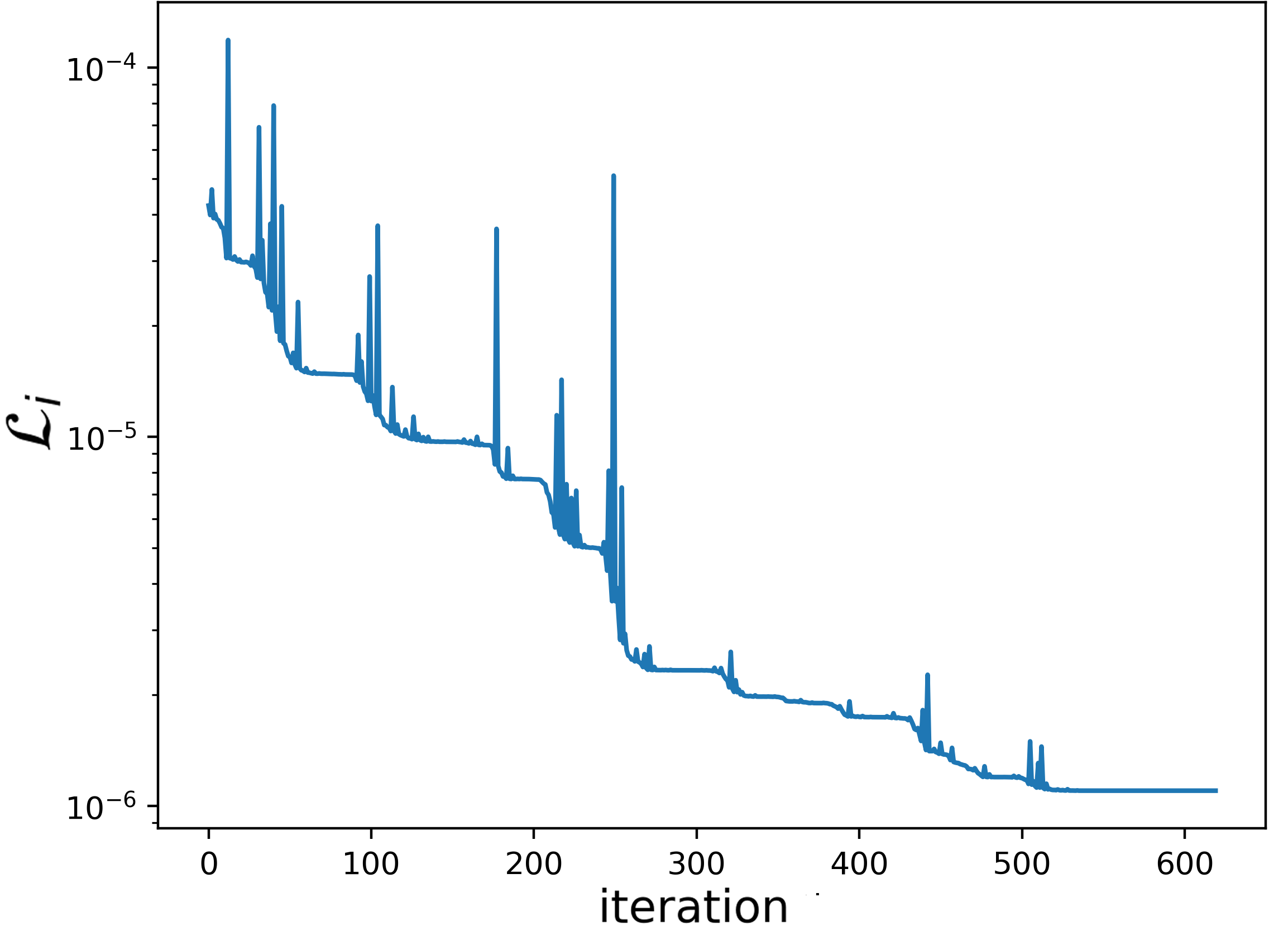}
        \caption{Convergence primal update \al iteration 1.}
        \label{Fig:conv_inner}
    \end{subfigure}

    \vspace{1em} 

    \begin{subfigure}[t]{0.475\textwidth}
        \centering
        \includegraphics[width=\linewidth]{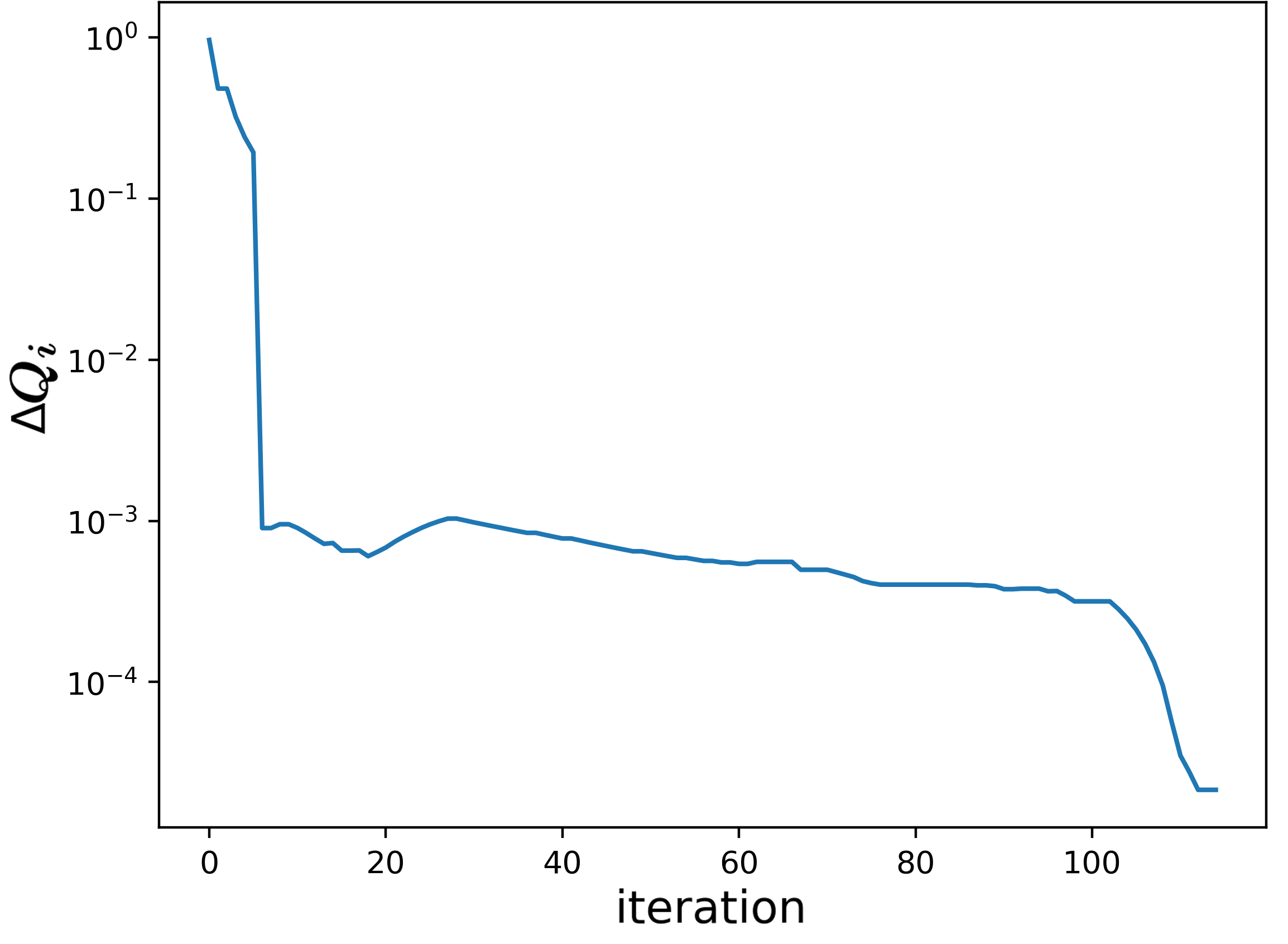}
        \caption{Convergence $\li$ loop \nl iteration 2.}
        \label{Fig:conv_outer_nl}
    \end{subfigure}
    \hfill
    \begin{subfigure}[t]{0.475\textwidth}
        \centering
        \includegraphics[width=\linewidth]{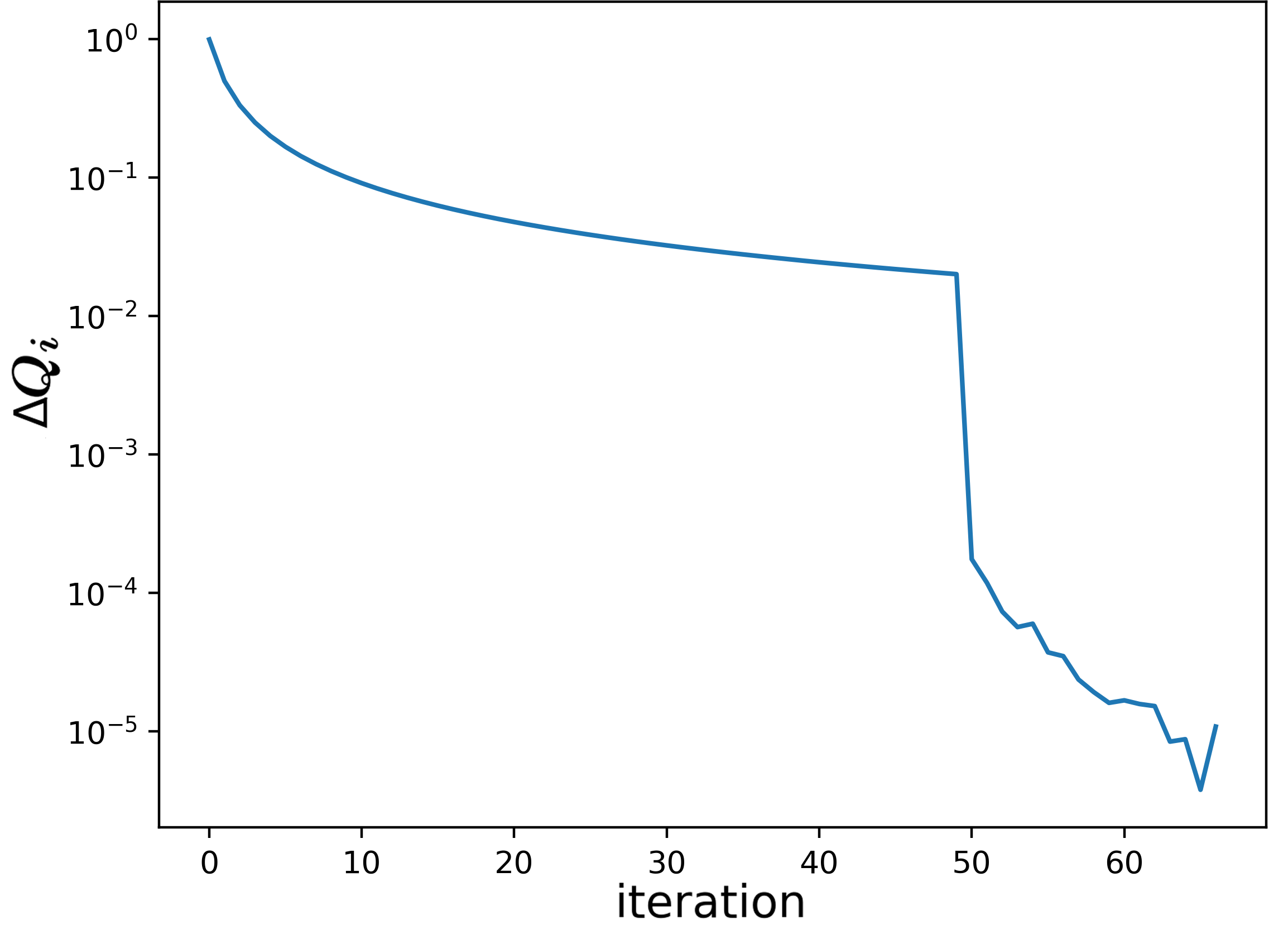}
        \caption{Convergence dual update \al iteration 1.}
        \label{Fig:conv_outer}
    \end{subfigure}

    \caption{Shows the convergence of the $\wi$ and dual updates of the \al and \nl methods. The subfigure captions indicate the iteration of the loop outside the given loops.}
    \label{Fig:conv_methods}
\end{figure}

To assess the convergence behavior of Alg.~\ref{Al:DA_nl} (\nl) and Alg.~\ref{Al:DA_al} (\al), the optimization loops for the network parameters $\wi$ and the Lagrange multipliers $\li$ are analyzed separately, which are denoted as dual and primal updates in the algorithms. 
These will be denoted as primal update and dual update.
For the test, the domain is split into three vertically stacked domains and 10 points are considered per interface.
The results are summarized in Fig.~\ref{Fig:conv_methods}, which consists of four subfigures: the top row shows the primal update gradient-descent convergence for \nl (left) and \al (right), while the bottom row displays the corresponding dual update gradient-ascent convergence. 
The horizontal axes indicate the number of iterations for each loop, and the vertical axes show the respective convergence measures.
For both algorithms, the second $\fm$ iteration is taken to show $\li$ convergence and the first $\li$ iteration is taken to show $\wi$ convergence.
Convergence of the primal update is evaluated by tracking the objective functions defined in Eqs.~(\ref{Eq:main_eq}, \ref{Eq:main_eq2}) for fixed values of $\lambda_i$. 
As shown in the top row of Fig.~\ref{Fig:conv_methods}, both formulations show a steady decrease in objective function values over successive iterations during a representative outer step (2). 
This consistent decrease, along with the diminishing slope, indicates stable convergence.
In \nl, a smoother convergence is observed, which can be attributed to the feasibility checks incorporated during linearization. 
These checks delay the introduction of potentially large penalty terms until constraints are satisfied, thereby stabilizing the loss evolution and resulting in more uniform optimization progress, albeit with fewer LBFGS steps per iteration.
Simultaneously, \al shows convergence to a lower value in the loss function, even if the loss value is generally higher for \al due to the additional sum.
Dual update convergence is assessed by monitoring the constraint approximations of the algorithm, according to Algs.~\ref{Al:DA_nl} and \ref{Al:DA_al}. 
Here, the change in $\Constr$ is used to show convergence, which is defined as: $\Delta \Constr=\mathbb{E}\left[\left|\Constr(\wi^{d})-\Constr(\wi^{d+1})\right|\right]$, instead of vanishing of the $\Constr$ itself, as the point of interest is where the $\Constr$ is not improving anymore.
As depicted in the bottom row of Fig.~\ref{Fig:conv_methods}, a clear trend towards zero in $\Delta \Constr$ is observed for both methods, indicating that the constraint residuals diminish consistently over outer iterations. 
In \al, smoother convergence is achieved due to the use of steepest gradient descent (SGD) \cite{NoceWrig06}, where the learning rate is scaled by the penalty parameter, appearing in Eq. \ref{Eq:main_eq2}. 
In contrast, \nl utilizes the Nadam optimizer (Nesterov-accelerated Adaptive Moment Estimation) \cite{nadams}, which allows for more adaptive learning rates and generally leads to faster convergence, though with slightly less regular behavior. 
For \al, a sharp drop in $\Delta \Constr$ occurs once the constraints approach zero. 
At this point, small oscillations around the feasibility threshold may cause the constraint signs to change, resulting in reversals in the direction of updates. 
In all cases, the optimization loops are terminated once the respective convergence tolerances are met.

\begin{figure}[t]
\centering
\includegraphics[width=0.2\textwidth]{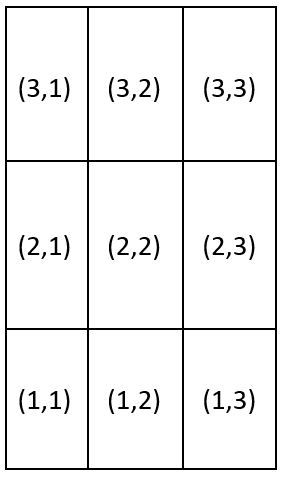}
\caption{Schematic of a (3,3) decomposition of a rectangular domain.} 
\label{Fig:cut_ex}
\end{figure}

The influence of the number of subdomains on performance of the two algorithms is assessed in the subsequent analysis. The test data is partitioned into $M = 2, 3, ..., 10$ domains, discarding points located on the interfaces to avoid conflicts in the local and interface objective.
Each local NN in this subsection is configured with identical hyper-parameters, as listed in Tab.~\ref{Tab:nn_params_2d}. Each subdomain can be identified by an index pair $(k, l)$, as illustrated in Fig.~\ref{Fig:cut_ex} for a 3x3 decomposition. 
Continuity is enforced by defining ten collocation points along each interface.
\begin{table}[t]
    \centering
    \caption{NN parameters 2D specimen for subdomain $(p,q)$.}
    \begin{tabular}{lrllllll}
    \toprule
        \multicolumn{1}{c}{Network} & \multicolumn{3}{c}{Architecture} & \multicolumn{1}{c}{Activation} & Optimizer & \multicolumn{1}{c}{Initializer} & penalty parameter\\ \cmidrule(r){2-4}
         \multicolumn{1}{c}{name} & \multicolumn{1}{c}{width} & x & \multicolumn{1}{c}{depth} & \multicolumn{1}{c}{function} &  & rate & (for \al)\\ \midrule
         $\hat\fvar_{\theta_{pq}}$ & 40 & x & 2 & swish & LBFGS & He Normal & $1 \cdot 10^-3$\\
    \bottomrule
    \end{tabular}
    \label{Tab:nn_params_2d}
\end{table}
\begin{figure}[t]
\centering
\begin{subfigure}[h]{0.475\textwidth}
    \centering
    \includegraphics[width=1.1\textwidth]{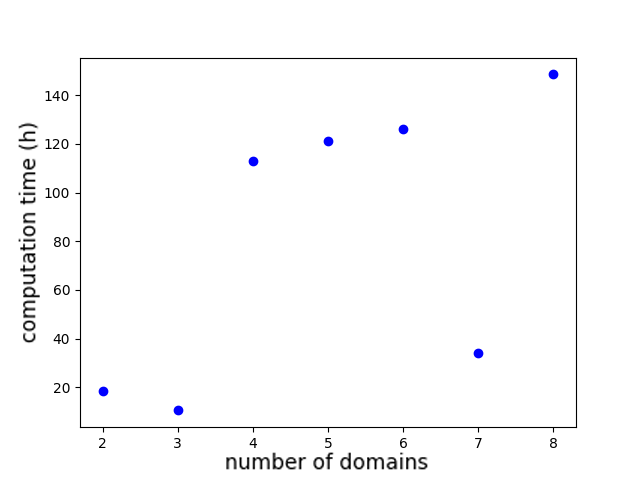}
    \caption{\nl.}    
    \label{Fig:ai_nl}
\end{subfigure}
\begin{subfigure}[h]{0.475\textwidth}  
    \centering 
    \includegraphics[width=1.1\textwidth]{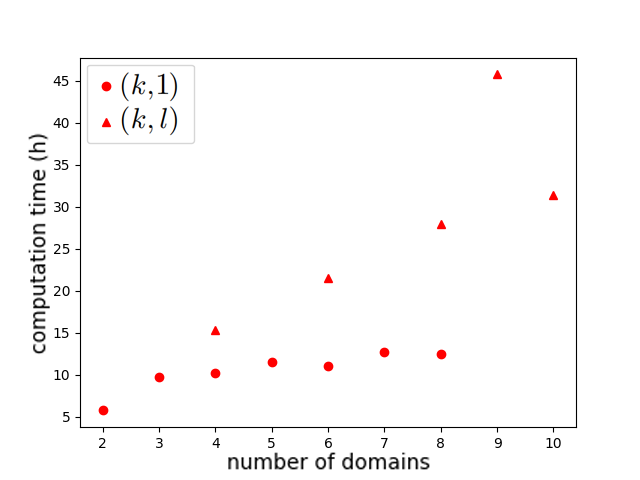}
    \caption{\al.}
    \label{Fig:ai_al}
\end{subfigure}
\caption{Full convergence times for different numbers of domains for the \nl and \al. The circles indicate domains split only horizontally, while the triangles are denoting runs for domain splits in both directions. For example for four domains, the circle is a run split into four by one domains and the triangle symbol run is two by two domains. The triangle symbol runs are left out in a) since they did not finish within the maximum time.} 
\label{Fig:ai_methods}
\end{figure}
Using this setup, both algorithms are used to train local NNs that approximate the normalized horizontal displacement field. The resulting computational costs and training behaviors are analyzed with respect to the number and arrangement of subdomains. Fig.~\ref{Fig:ai_methods} summarizes the total computational time over 100 training iterations as a function of number of domains, in which 10 points are considered per interface. Detailed results for the \nl and \al methods are shown in Figs.~\ref{Fig:ai_nl} and \ref{Fig:ai_al}, respectively. Circular markers indicate vertical-only splits, i.e., configurations of the form $(M,1)$, whereas triangular markers represent multi-directional splits with $M = K \cdot L$ domains (e.g., the configuration shown in Fig.~\ref{Fig:points} corresponds to $(3,1)$).
For the \nl formulation, the lowest total parallel training time ($10.4$ h) is observed when the domain is divided into three vertical subdomains. In contrast, the \al formulation achieves its minimal training time ($5.8$ h) with only two domains. When increasing the number of subdomains, \nl shows a large increase in computational cost, whereas \al displays a nearly linear growth, particularly for vertical splits. Multi-directional splits result in fast growth of computational times for both methods, though only the \al stays within reasonable computational times.
The large difference in computational times observed in \nl is attributed to the algorithm’s sensitivity to the underlying physics of the problem. In particular, it is expected that the boundary deformation region near the bottom of the domain introduces training difficulty when split into excessively small subdomains. In such cases, the network must approximate high gradients within a narrow region, leading to long convergence times. An exception to this trend appears at $M = 7$, where the interface happens to align with the transition between linear regime and boundaries. Hence, the lowest NN only has to approximate the boundary part, while the others approximate the linear parts. Meaning that it is beneficial to train the boundary separately from the rest of the domain.

When comparing the total simulation times of the performed simulations, it is noticeable that the \al generally shows faster convergence than the \nl across different domain decompositions. Computational times for the \nl range between $10$ and $140$ hours, while the \al remains between $5$ and $45$ hours, indicating superior scalability of the \al for increasing numbers of subdomains. The scalability issue of \nl becomes even more severe looking at multi-directional splits of \nl, which exceeded the maximum allowed training time of $168$ hours.
\begin{figure}[t!]
    \centering
    \begin{subfigure}[b]{0.475\textwidth}
        \centering
        \includegraphics[width=1.1\textwidth]{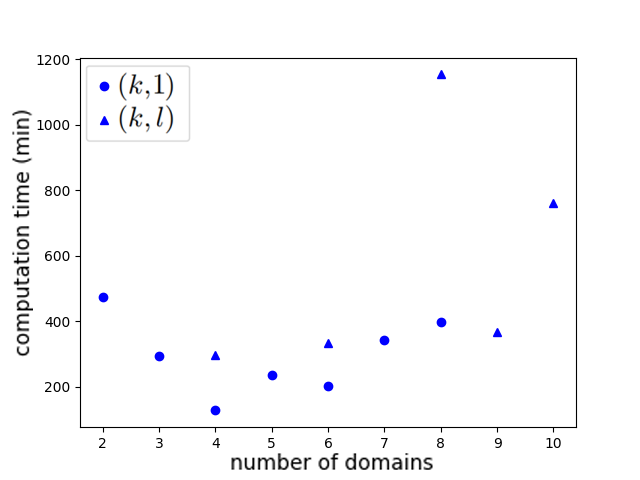}
        \caption{\nl.}    
        \label{Fig:si_nl}
    \end{subfigure}
    \begin{subfigure}[b]{0.475\textwidth}  
        \centering 
        \includegraphics[width=1.1\textwidth]{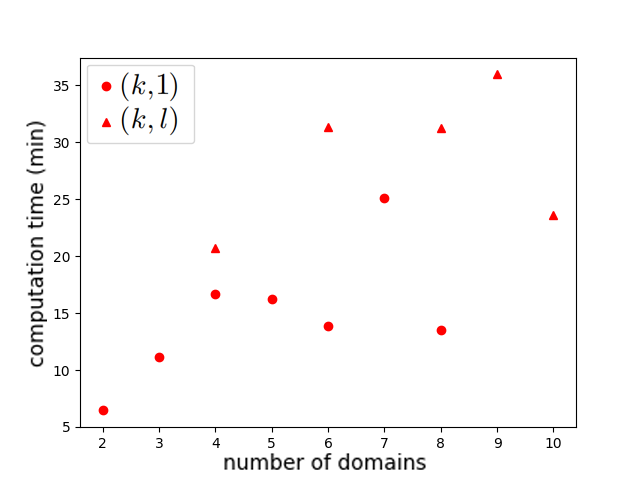}
        \caption{\al.}
        \label{Fig:si_al}
    \end{subfigure}
    \caption{Convergence times of a single iteration for different numbers of domains for the \nl and \al. The circles indicate domains split only horizontally, while the triangles are denoting runs for domain splits in both directions. For example for four domains, the circle is a run split into four by one domains and the triangle symbol run is two by two domains.} 
    \label{Fig:si_methods}
\end{figure}
\begin{figure}[b!]
    \centering
    \includegraphics[width=0.522\textwidth]{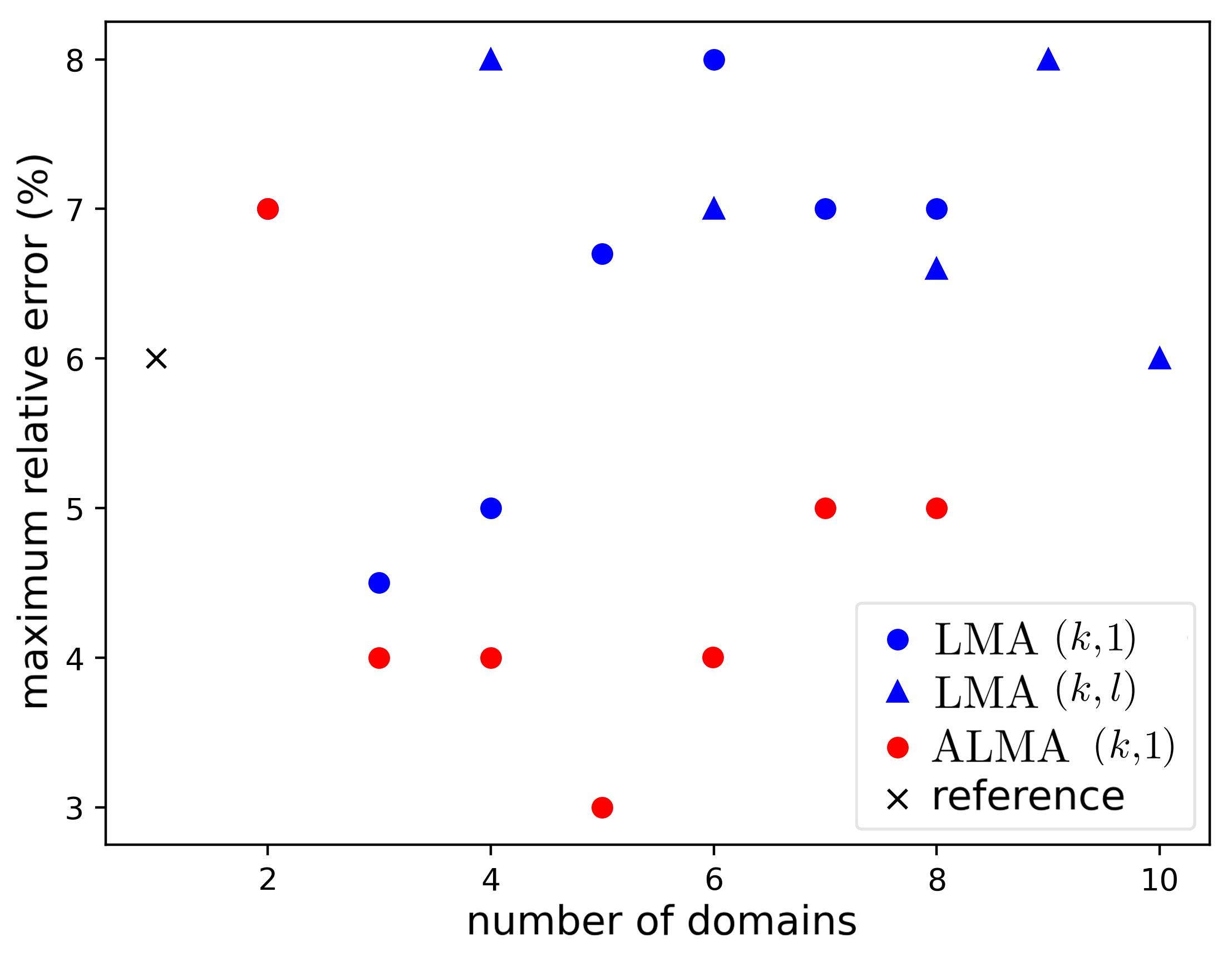}
    \caption{Final relative errors for different numbers of domains for the \nl and \al. The circles indicate one directional splits, while the triangles denote multi-directional splits. Multi-directional splits of the \al are left out since they have higher relative errors. The black x denotes a reference solution for a single NN trained on the whole domain.} 
    \label{Fig:rel_max_d}
\end{figure}
Further insights into the computational efficiency of the two methods are provided by Fig.~\ref{Fig:si_methods}, which shows the computational time for a representative iteration (iteration 2) across different domain decompositions. This iteration is typically one of the most computationally intensive, as it is after the first update of the Lagrange multipliers and often requires repeated linearization steps to reach a feasible step size. Note, that the search for a feasible step size is not required for the \al formulation. For the \nl, iteration times typically range between 200 and 400 minutes, with relatively little variation across different number of subdomains. In contrast, the \al demonstrates substantially shorter iteration times, ranging from approximately 10 to 35 minutes, depending on the number and configuration of subdomains.
Interestingly, while the computational cost per iteration of the \al increases with the number of subdomains—particularly for multi-directional splits, this growth cannot be seen in the \nl with consistently high cost. The relatively consistent computational times of the \nl are attributed to their already intensive per-iteration workload, due to repeated linearization and constraint feasibility checks. Essentially, the method takes an update step in one direction and discards it afterwards, if the point is not feasible. Consequently, the number of domains has limited additional impact on its performance, whereas the \al, being more light-weight in each iteration, exhibits a clearer dependency on the number of interface conditions introduced by the domain decomposition.

To assess the accuracy of both methods, a maximum relative error is defined from the same simulations as previously stated.
This relative error is slightly modified to account for zero values in the FEM points $\fvar_i(x)$. Fig.~\ref{Fig:rel_max_d} presents the maximum relative error - computed according to:
\begin{figure}[b!]
    \centering

    \begin{subfigure}[t]{0.3\textwidth}
        \centering
        \includegraphics[width=\linewidth]{Z-real.png}
        \caption{Measurement solution.}
    \end{subfigure}
    \hfill
    \begin{subfigure}[t]{0.3\textwidth}
        \centering
        \includegraphics[width=\linewidth]{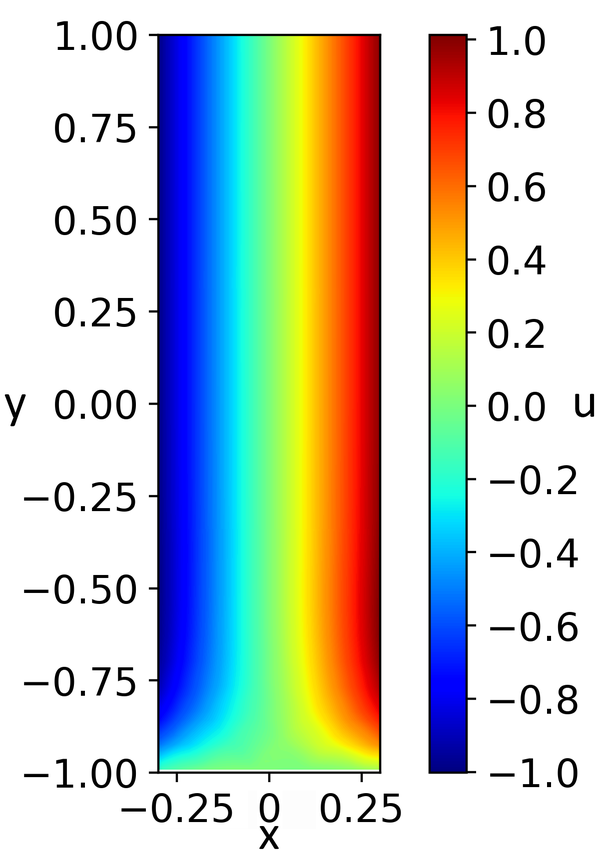}
        \caption{\nl solution.}
    \end{subfigure}
    \hfill
    \begin{subfigure}[t]{0.3\textwidth}
        \centering
        \includegraphics[width=\linewidth]{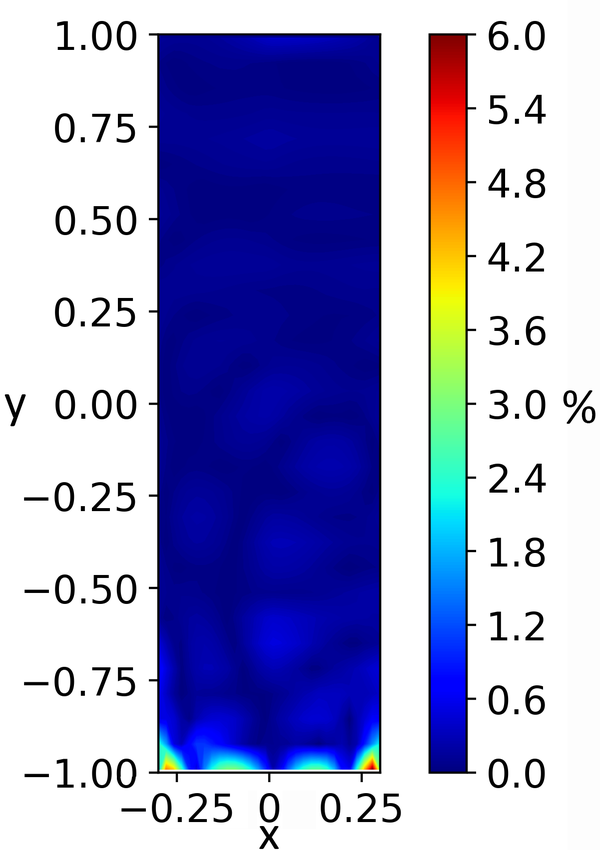}
        \caption{Relative error single domain.}
    \end{subfigure}

    \vspace{1em} 

    \begin{subfigure}[t]{0.3\textwidth}
        \centering
        \includegraphics[width=\linewidth]{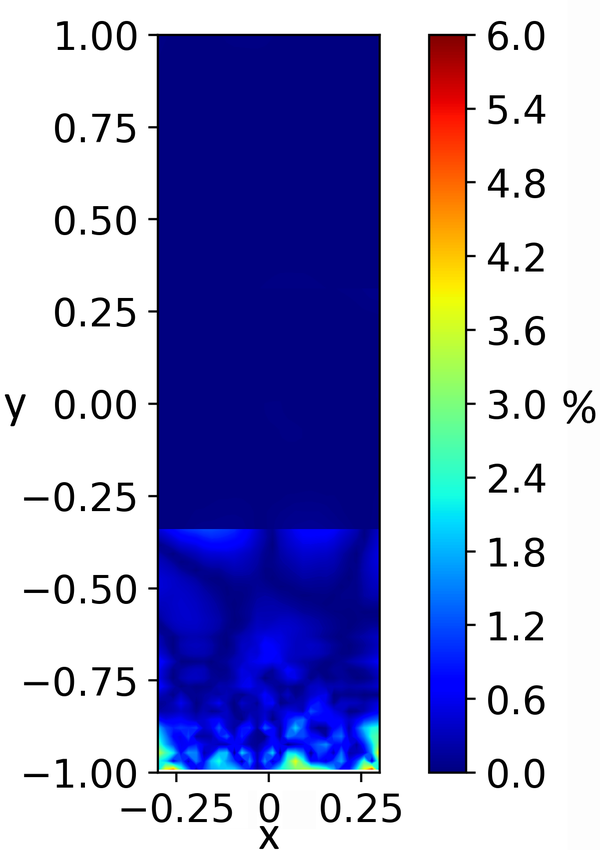}
        \caption{Relative error without continuity constraints.}
    \end{subfigure}
    \hfill
    \begin{subfigure}[t]{0.3\textwidth}
        \centering
        \includegraphics[width=\linewidth]{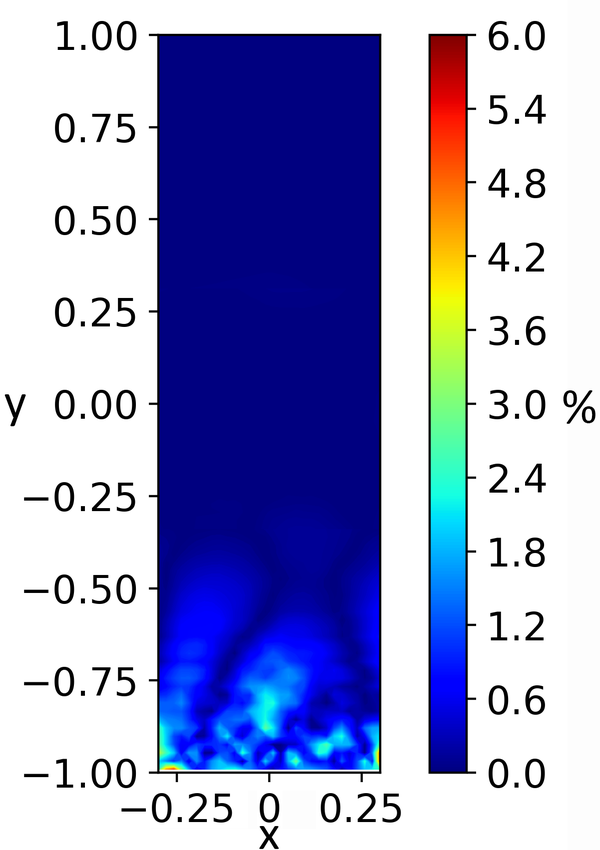}
        \caption{Relative error \nl.}
    \end{subfigure}
    \hfill
    \begin{subfigure}[t]{0.3\textwidth}
        \centering
        \includegraphics[width=\linewidth]{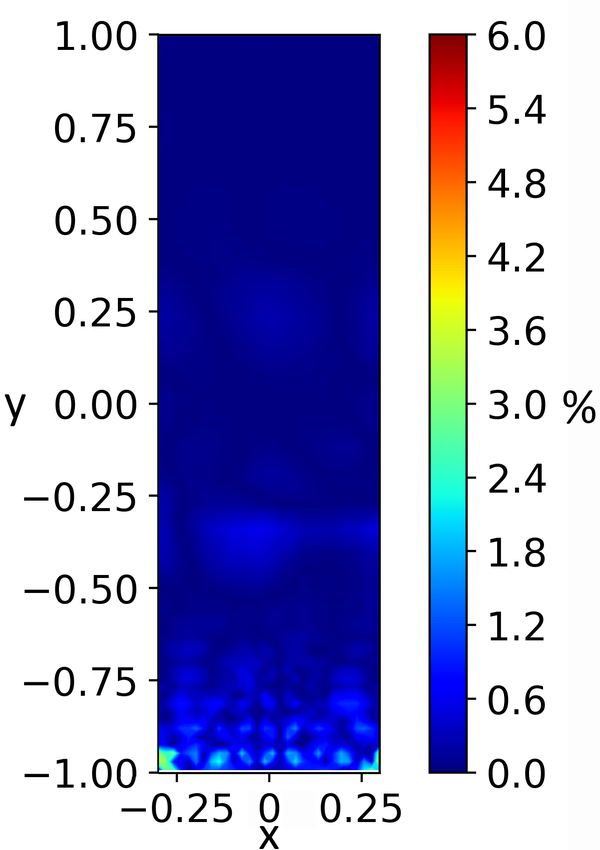}
        \caption{Relative error \al.}
    \end{subfigure}

    \caption{FEM solution and approximations of the solution split into three subdomains. One can view the FEM solution in a), the \al and \nl approximations in b) and c), the relative error of each of these methods in e) and f), and the relative error without continuity constraints enforced in d).}
    \label{fig:three_domains_approx}
\end{figure}
\begin{equation}
    e_{rel} = \frac{|\fwi(x) - \fvar_i(x)|}{|\fvar_i(x)| + 1},
    \label{Eq:rele+1}
\end{equation}
\begin{figure}[b!]
    \centering

    \begin{subfigure}[t]{0.3\textwidth}
        \centering
        \includegraphics[width=\linewidth]{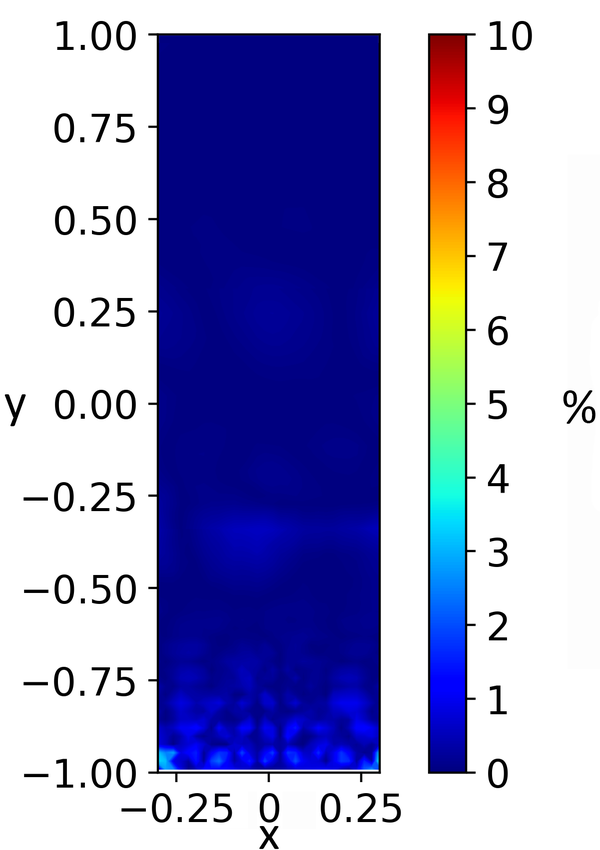}
        \caption{Gap size of one row ($\approx 0.033$).}
    \end{subfigure}
    \hfill
    \begin{subfigure}[t]{0.3\textwidth}
        \centering
        \includegraphics[width=\linewidth]{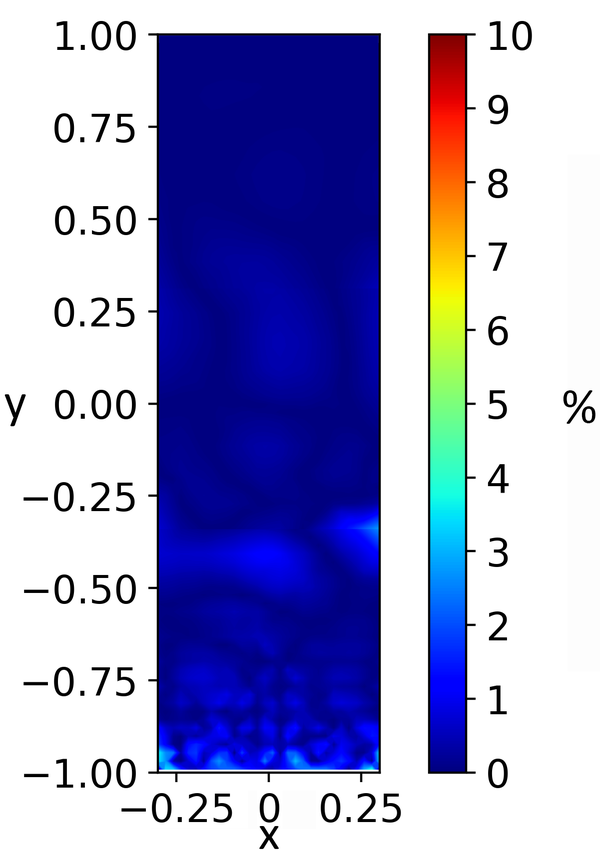}
        \caption{Gap size of two rows ($\approx 0.1$).}
    \end{subfigure}
    \hfill
    \begin{subfigure}[t]{0.3\textwidth}
        \centering
        \includegraphics[width=\linewidth]{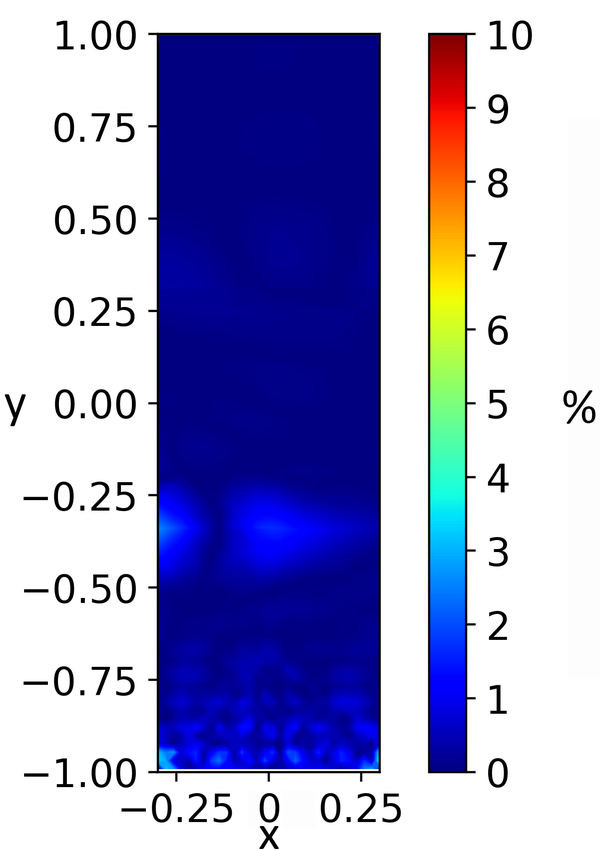}
        \caption{Gap size of three rows ($\approx 0.166$).}
    \end{subfigure}

    \vspace{1em} 

    \begin{subfigure}[t]{0.3\textwidth}
        \centering
        \includegraphics[width=\linewidth]{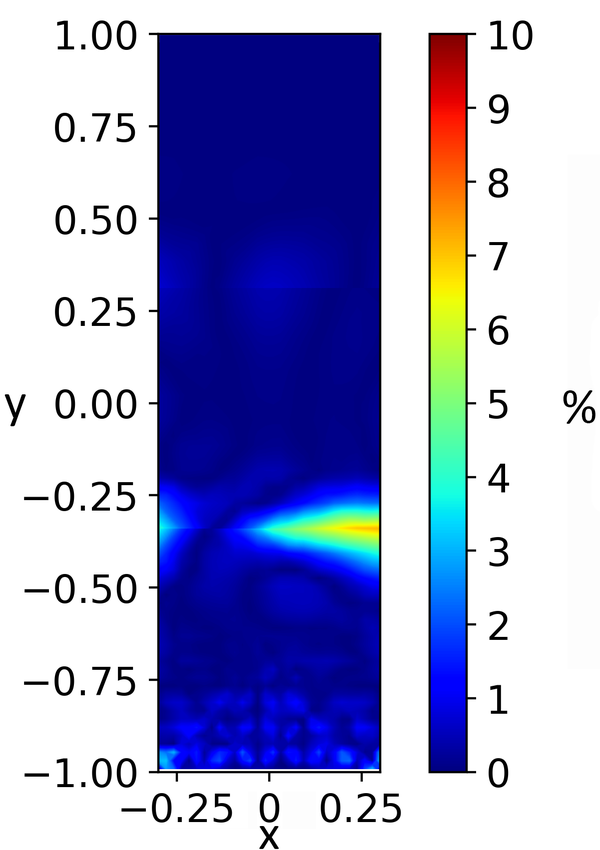}
        \caption{Gap size of four rows ($\approx 0.233$).}
    \end{subfigure}
    \hfill
    \begin{subfigure}[t]{0.3\textwidth}
        \centering
        \includegraphics[width=\linewidth]{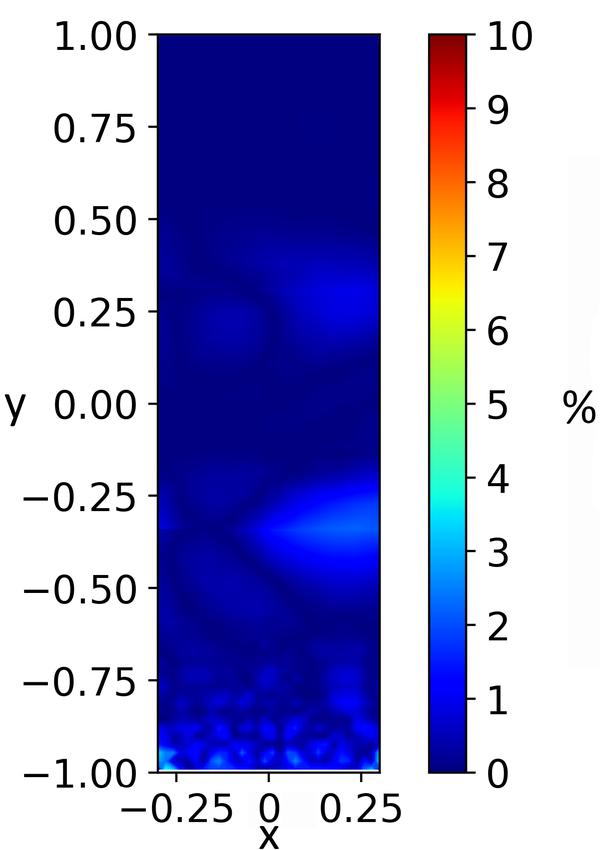}
        \caption{Gap size of five rows ($\approx 0.3$).}
    \end{subfigure}
    \hfill
    \begin{subfigure}[t]{0.3\textwidth}
        \centering
        \includegraphics[width=\linewidth]{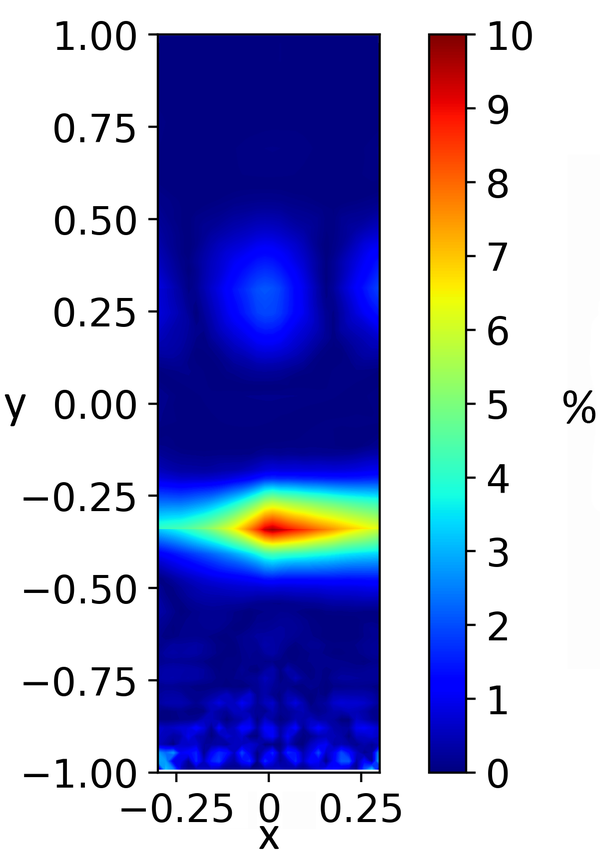}
        \caption{Gap size of six rows ($\approx 0.366$).}
    \end{subfigure}

    \caption{Relative errors for \al approximations with different gap sizes between three vertical domains in which the interface constraints are enforced halfway across the gaps.}
    \label{fig:td_gapsize}
\end{figure}
- against the number of subdomains. The shift by $1$ in the denominator omits division by zero since the values are normalized to be bounded by $-1$ and $1$. 
Note, that the multi-directional splits of \al are excluded due to final maximum relative errors exceeding $30\%$ and the \nl results from the last completed iteration are shown since they take more than the maximum training time to compute. For the \nl results that means not all $\fm$ updates are performed, while the $\li$ and $\wi$ are converged. The multi-directional decompositions are shown as triangles, while the vertically stacked decompositions are shown as circles. Lagrange maximums relative errors are shown in blue, while augmented Lagrange errors are shown in red. A reference solution of the maximum relative error of a single NN with $80$ x $2$ neurons is shown as a black cross.
When compared to the \nl, the maximum relative error of \al shows the higher errors near the subdomain interfaces, suggesting that the \al tends to approximate these less accurately. The interface error occurs due to an accumulation of errors since the \al accepts the approximation faster than the more exact \nl. However, this improvement in continuity, i.e., precision, for the \nl comes at the cost of computational time, as it requires multiple attempts to determine a feasible Lagrange multiplier update suitable for a functioning linearization. In contrast, the \al introduces a tunable penalty parameter that can be adjusted to mitigate interface errors in a more flexible way.
The minimum overall relative error is achieved by the \al when the domain is decomposed into five subdomains. Whereas comparable accuracy is observed for configurations with three to eight subdomains. When both computational time and approximation accuracy are considered jointly, a decomposition into three vertical subdomains using the \al provides an optimal balance between runtime and accuracy.
Fig.~\ref{fig:three_domains_approx} presents the results for this configuration. Subplots (a)–(b) display the measured solution, and the corresponding \nl approximations. Visual comparison indicates that the surrogate model closely reflect the reference solution, this is the same for \al. Subplots (c)–(f) show the corresponding relative errors: in the first row the relative error for a single domain, in the second row, first, for an unconstrained approximation (no continuity enforcement), followed by the results obtained using \nl and \al, respectively. 
The single-domain NN is trained within only $3.2$ min, which significantly faster than the presented algorithms. This is expected, as it relies on standard TensorFlow routines applied to this relatively simple problem while the proposed methods are presented for larger problems and higher accuracy through local approximations. 
However, the relative error plots confirm a higher accuracy for local approximations for the problem as shown in Fig.~\ref{Fig:rel_max_d}. 
The best results from DDMs achieve error levels approximately half those of the single-domain approach. 
It is also clearly visible that the errors near the bottom boundary are greatly reduced when comparing the single-domain approximation to the \nl and \al approximation.
Notably, continuity improves significantly between the lower subdomains when the continuity constraints are enforced by \nl and \al. The \nl reduces discontinuities at the cost of some loss of accuracy in the lower regions, whereas the \al improves both continuity and local accuracy compared to the unconstrained model.
Based on these findings, the \al with three vertically stacked subdomains is used for further investigations, focusing on the sensitivity of the method to other algorithmic parameters.

In many engineering problems, one deals with missing data.
In particular, in applications in which only part of the QoI can be sensed, such that no full-field measurements are available
This limitation is particularly prominent in digital image correlation, in which a speckle pattern is applied to a surface of a specimen and tracked using cameras to obtain displacement and strain fields. 
Therefore, in this section the performance of NN DDM on this type of problem is analyzed by artificially introducing a lack of data.
As in the previous examples, the NN parameters for each local NN are chosen as in Tab.~\ref{Tab:nn_params_2d}.
In doing so, an increasing number of lines of data points around the interfaces is discarded. 
Discarding points is done in a line by line manner, as can be seen in Fig.~\ref{Fig:points}, in which the round black points represent an interface. 
To introduce the lack of data near the interface, the blue triangle and red square points from the local domains that are the closest to the interface are discarded, the closer the points are to the interface.
In  Fig.~\ref{fig:td_gapsize}, the relative errors after convergence are shown for the gap size increased from one row ($\approx 0.033$) to six rows of data ($\approx 0.366$) from Fig.~\ref{fig:td_gapsize} a to f. 
Here, it can be seen that the overall maximum relative error is not largely influenced by increasing the size of the gap between domains. 
But of course, in creating larger gaps between local domains, the extrapolation capabilities of the NNs predicting the local domains are tested. 
Since NNs are bad at extrapolating, the NNs have bad approximations when predicting outside the data points.
This leads to a bad approximation at the interface from both domains.
Since the local approximations are the training points for the interface approximations $\fm$, also this approximation becomes inaccurate as no data is known on the interfaces.
There are infinitely many solutions for the interface lines due to the larger area with lack of data around the interfaces.
Hence, every run of the algorithm will lead to a different approximation. 
Therefore, one cannot really gain information from a higher maximum relative error for this single discrete converged approximation at the lower interface of Fig.~\ref{fig:td_gapsize}d) as the approximation is inherently stochastic with lack of data around the interface.

\begin{figure}[b!]
    \centering

    \begin{subfigure}[t]{0.24\textwidth}
        \centering
        \includegraphics[width=\linewidth]{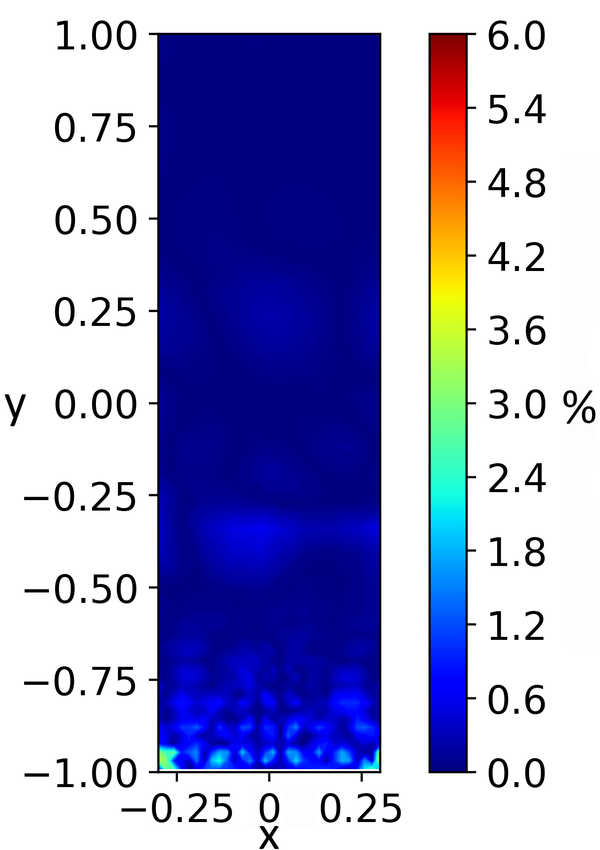}
        \caption{$1356$ points per domain.}
    \end{subfigure}
    \hfill
    \begin{subfigure}[t]{0.24\textwidth}
        \centering
        \includegraphics[width=\linewidth]{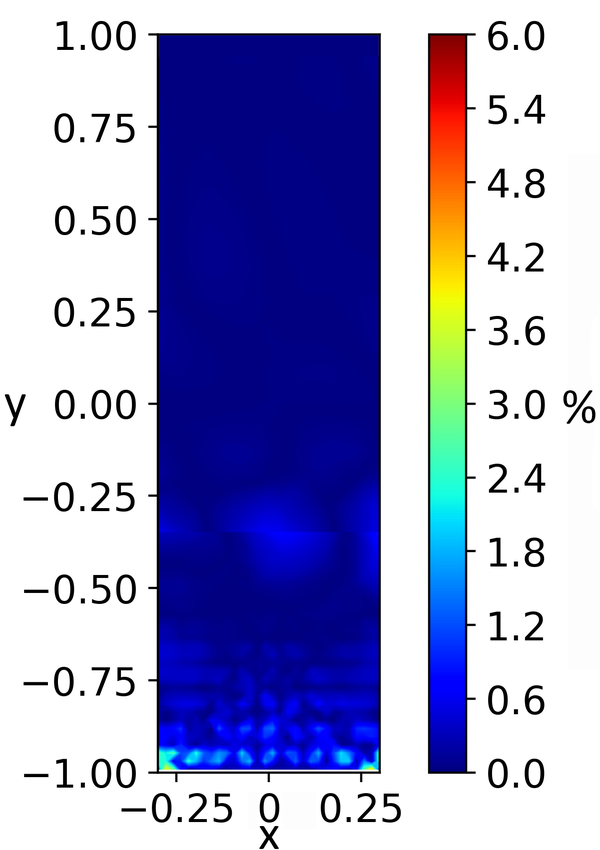}
        \caption{$617$ points per domain.}
    \end{subfigure}
    \hfill
    \begin{subfigure}[t]{0.24\textwidth}
        \centering
        \includegraphics[width=\linewidth]{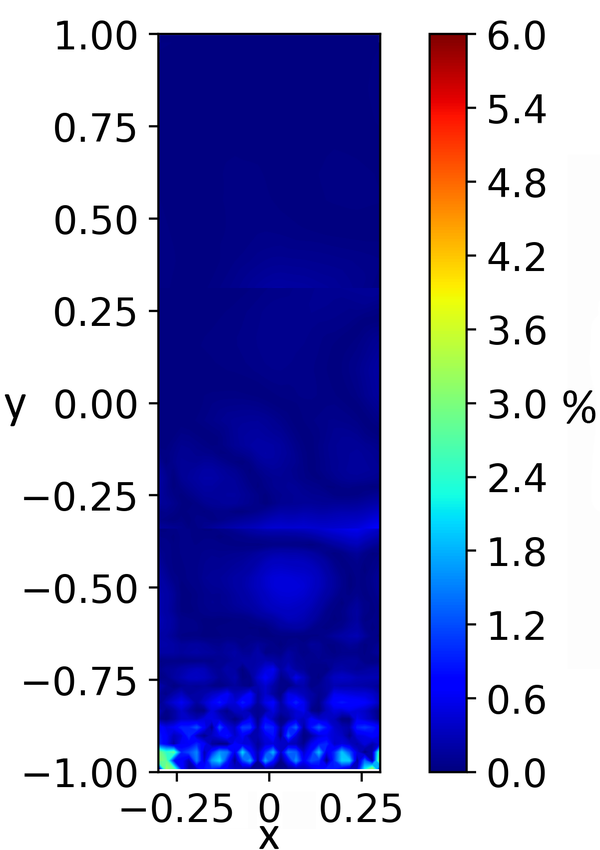}
        \caption{$361$ points per domain.}
    \end{subfigure}
    \hfill
    \begin{subfigure}[t]{0.24\textwidth}
        \centering
        \includegraphics[width=\linewidth]{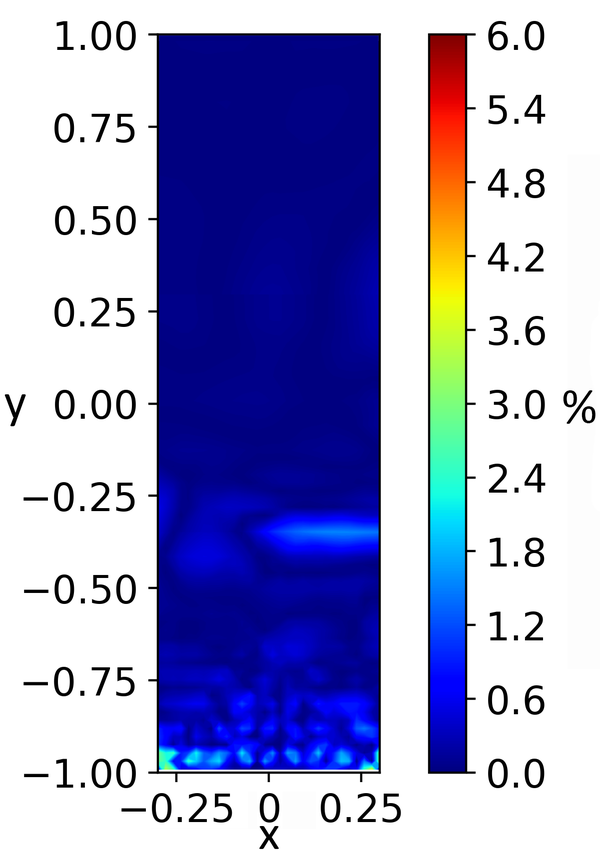}
        \caption{$225$ points per domain.}
    \end{subfigure}
    \caption{Relative errors for three vertically stacked domains for \al. The number of points per local domain decrease from a) to d) with the exact number of points per dom as denoted above.}
    \label{Fig:more_points}
\end{figure}

The number of training points per domain are varied to analyze its influence on the \al, as can be seen from Fig.~\ref{Fig:more_points}. For these simulations, the local NN properties remain according to Tab.~\ref{Tab:nn_params_2d}.
Here, relative errors for the three domains are compared when the number of points per the domains decreases (from left to right). 
In terms of computational times, both algorithms equally benefit from a lower number of points per domain.
In contrast to expectation, the maximum relative error is observed to decrease as the measurement points decrease.
This counterintuitive behavior can be explained by the local high gradients at the bottom of the domain. 
When the number of points per domain is low, the boundary is not captured well.
That means the number of data points near the boundary is not large enough to capture the behavior near the boundary based on the data.
Since the points are ordered in a grid, the next line of points is further away from the boundary when the number of points is reduced.
This leads to a larger relative error in a larger area near the boundary when compared to using a larger number of points per domain. 
Given a larger number of points per domain, the boundary is approximated well, which moves the highest error to the lower edge of the figure. 
In addition to that, the lower part of Fig.~\ref{Fig:more_points} with the lowest number of points shows that the error along the interface increases. 
This can be attributed to the coarser grid introducing a larger distance between the local domains, similar to the effect of increasing the gap size at the interface in the previous test.

Another benefit of having locally defined NNs is that one can use different architectures for different parts of the domain to reduce the total number of required parameters. 
Using a low number of parameters at the top to approximate the almost linear part while having more parameters to train the bottom. 
This will not change the results or computational times, since the lowest NN is still the part the has the largest computational times and approximation errors. 
However, it is possible to set memory free faster since the other parallel computed domains already converged. 
Finally, the additional memory can be used to faster compute the most expensive computation at the lowest subdomain of the problem or to do other computations.

\subsection{Uncertainty propagation in 3D cylinder for optimization}
\label{Sec:3D}
\begin{figure}[t]
    \begin{subfigure}[b]{0.3\textwidth}  
    \centering 
    \includegraphics[width=0.6\textwidth]{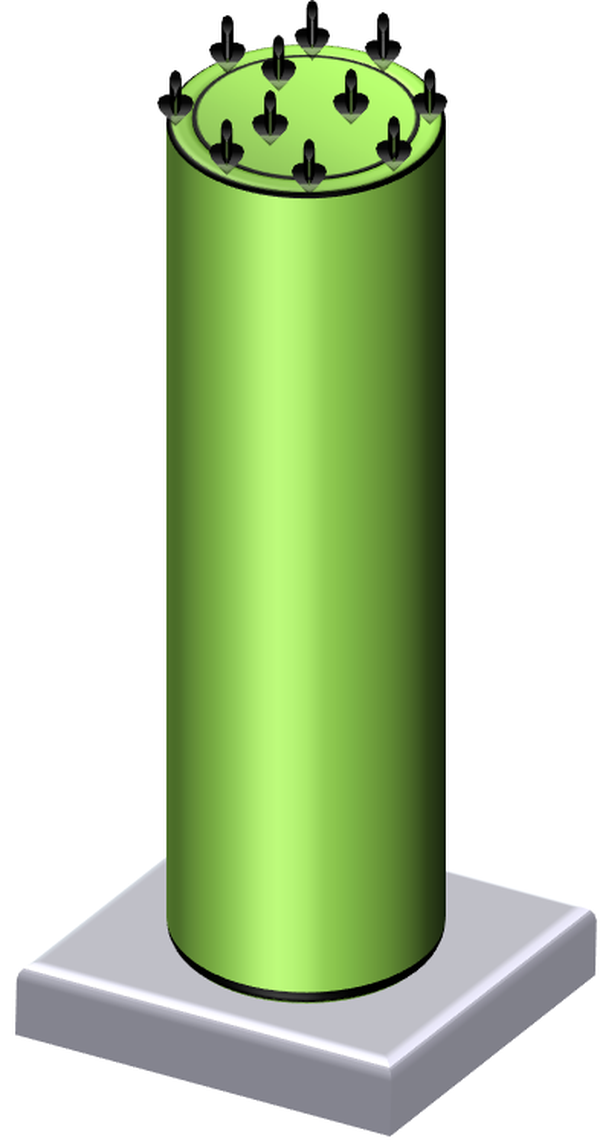}
    \caption{}
    \label{Fig:3dcyl2}
    \end{subfigure}
    \begin{subfigure}[b]{0.3\textwidth} 
    \centering 
    \includegraphics[width=0.65\textwidth]{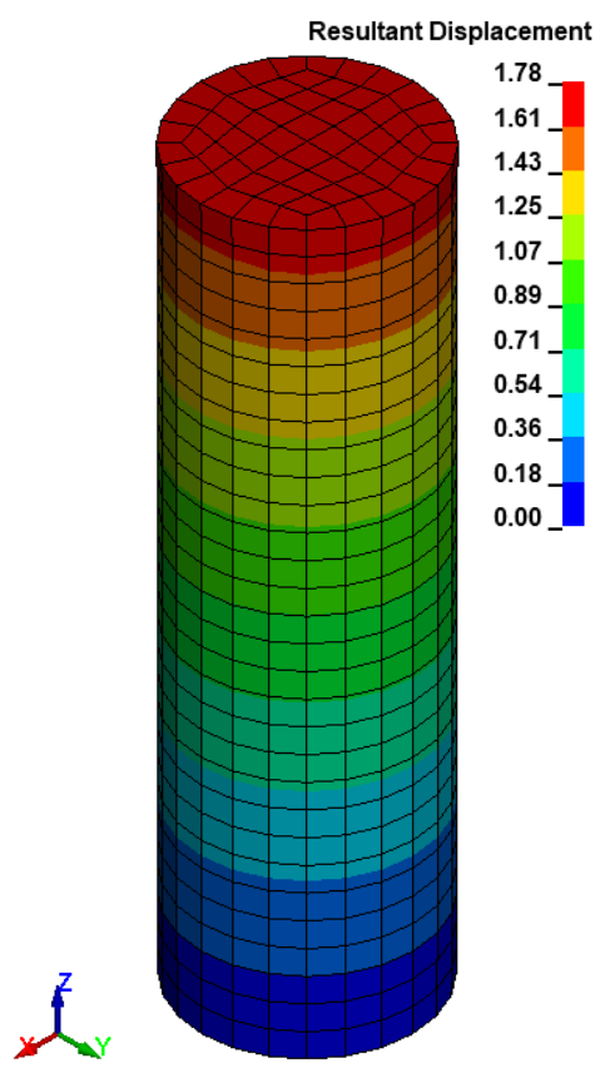}
    \caption{}
    \label{Fig:3dcyl3}
     \end{subfigure}
    \begin{subfigure}[b]{0.3\textwidth} 
    \centering 
    \includegraphics[width=0.4\textwidth]{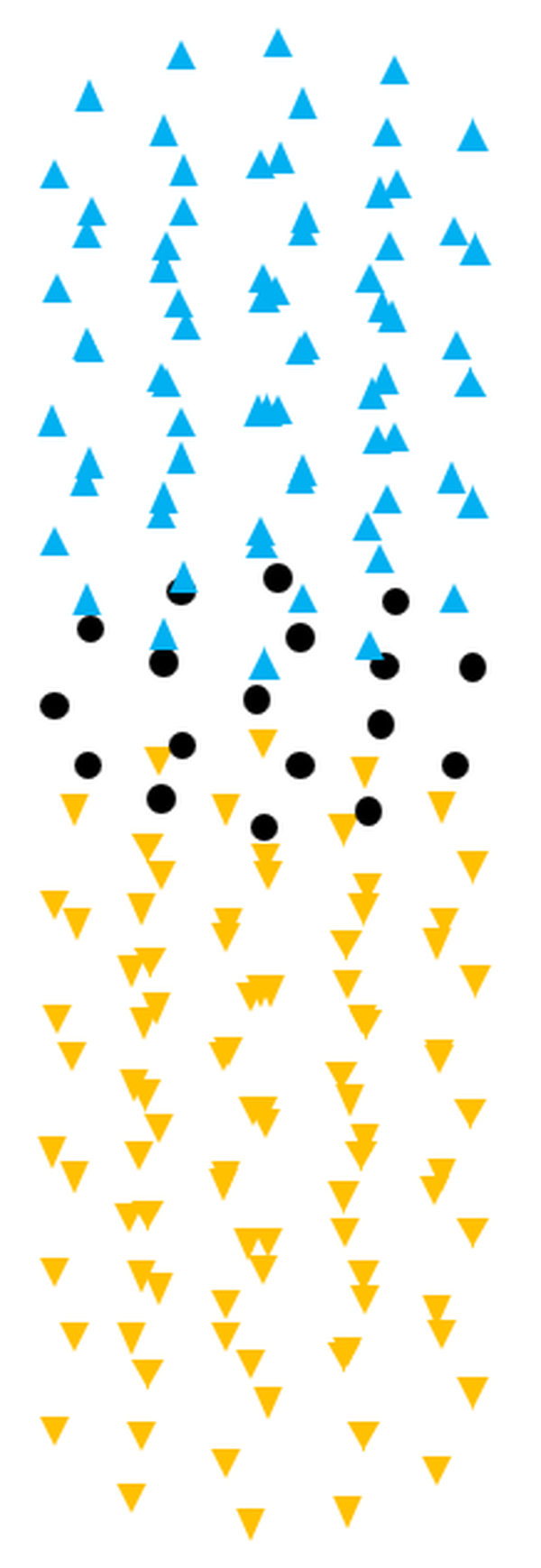}
    \caption{}
    \label{Fig:3dcyl}
     \end{subfigure}
    \caption{a) Schematic of the 3D Cylinder clamped at the bottom and compressed at the top. b) Resultant displacement field of the FEM simulation for one random set of material parameters. c) Data points on the cylinder split into two domains with the upper domain in blue, the lower domain in yellow and the interface in black.} 
    \label{Fig:problem}
\end{figure}
A material parameter optimization problem is considered to demonstrate the applicability of the NN DDM surrogate modeling framework.
The objective is to make a surrogate model that can be used to minimize the displacement of a three-dimensional, linearly-elastic cylinder subjected to a compressive load of $20$ kN, see Fig. \ref{Fig:3dcyl2}. 
The cylinder has a diameter of $20$ mm and a height of $70$ mm and is clamped at the bottom and compressed at the top surface. 
Material parameters, specifically the bulk modulus $\kappa$ and shear modulus $\mu$, are treated as optimization variables $\zeta(\omega) = [\kappa(\omega), \mu(\omega)]^T$, and modeled in a probabilistic setting, according to Eq. \ref{Eq:PDE3}. 
The moduli are modeled as lognormal random variables with means and standard deviations as shown in Tab.~\ref{Tab:3dcy}.
Their corresponding Monte Carlo samples are simulated in FEM to compute the corresponding displacement fields, as illustrated in Fig.~\ref{Fig:3dcyl3}.
The displacement field is defined in three components, such that $\fvar=[u_x, u_y, u_z]^T$.
Since the simulation is expensive, a surrogate model is needed to perform the simulation in a fast manner.
The surrogate model is constructed using the NN DDM with two spatial subdomains representing the upper and lower halves of the cylinder, as shown in Fig.~\ref{Fig:3dcyl}. 
That means the model is split in the spatial domain, while the stochastic domain of \(\zeta(\omega)\) remains global. 
The local domains have all probabilistic samples as input and only the upper and lower part of the spatial domain to generate a first-order smooth interface between the latter two spatial sets. 
For fast and highly precise network parameter updates of local domains, a LBFGS gradient-based optimizer is used, which utilizes second-order gradient approximations.
Hyper-parameters for the subdomains are listed in Tab.~\ref{Tab:nn_params_3d}. 
To ensure interface consistency between subdomains, interface constraints are enforced by \al across $73$ spatial points on the section surface for each of the $1000$ material parameter samples. 

\begin{table}[h]
    \centering
    \caption{Cylinder properties for the finite element simulations.}
    \begin{tabular}{llllllllllll}
    \toprule
        \multicolumn{3}{c}{Dimensions [mm$^3$]} & \multicolumn{2}{c}{Bulk mod. [GPa]} & \multicolumn{2}{c}{Shear mod. [GPa]} & \multicolumn{1}{c}{Element} & \multicolumn{2}{c}{Number of}\\
        \cmidrule(r){1-3} \cmidrule(r){4-5} \cmidrule(r){6-7} \cmidrule(r){9-10}
         \multicolumn{1}{c}{diameter} & x & \multicolumn{1}{c}{height} & \multicolumn{1}{c}{mean} & \multicolumn{1}{c}{variance} & \multicolumn{1}{c}{mean} & \multicolumn{1}{c}{variance} & \multicolumn{1}{c}{type} & \multicolumn{1}{c}{elements} &\multicolumn{1}{c}{nodes}\\
        \midrule
        20 & x & 70 & 175 & 10 & 81 & 10 & hexahedron & 1800 & 31 x 73\\
    \bottomrule
    \end{tabular}
    
    \label{Tab:3dcy}
\end{table}

\begin{table}[t]
    \centering
    \caption{Neural network parameters 3D cylinder for subdomain $(k,l,m)$.}
    \begin{tabular}{lrllllrr}
    \toprule
        \multicolumn{1}{c}{Network} & \multicolumn{3}{c}{Architecture} & \multicolumn{1}{c}{Activation} & Optimizer & \multicolumn{1}{c}{Initializer}\\ \cmidrule(r){2-4}
         \multicolumn{1}{c}{name} & \multicolumn{1}{c}{width} & x & \multicolumn{1}{c}{depth} & \multicolumn{1}{c}{function} &  &\\ \midrule
         $\hat\fvar_{\theta_{klm}}$ & 80 & x & 4 & swish & LBFGS & He Normal\\
    \bottomrule
    \end{tabular}
    \label{Tab:nn_params_3d}
\end{table}

\begin{figure}[h]
    \begin{subfigure}[b]{0.5\textwidth}  
    \centering 
    \includegraphics[width=\textwidth]{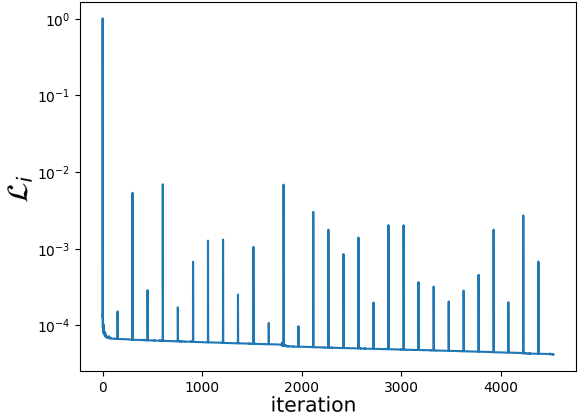}
    \caption{Convergence primal update \al iteration 4.}
    \label{Fig:conv_inner_3d}
    \end{subfigure}
    \begin{subfigure}[b]{0.5\textwidth} 
    \centering 
    \includegraphics[width=\textwidth]{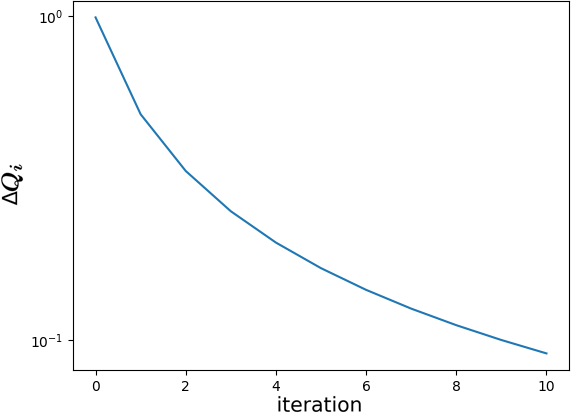}
    \caption{Convergence dual update \al iteration 2.}
    \label{Fig:conv_outer_3d}
     \end{subfigure}
    \caption{Shows the convergence of the $\wi$ and dual updates of \al for the three-dimensional problem. The subfigure captions indicate the iteration of the loop outside the given loops.} 
    \label{Fig:3D_conv}
\end{figure}

Initially, the convergence of the three-dimensional model is shown in Fig.~\ref{Fig:3D_conv}. 
Similar to in the previous example in Section \ref{Sec:2D}, the figure shows convergence of \al in network parameters $\wi$ and Lagrange multipliers $\li$. 
Both figures indicate the loop iterations on the x-axis and the corresponding convergence criterion of the loop from Alg.~\ref{Al:DA_al} show dual update convergence, while the fourth dual update is used to show $\wi$ convergence.
In case of the primal update, the criterion is the change in $\Lagr_i$, while the criterion for the dual update is a mean-absolute change $\Delta \Constr$.
In the primal update, several peaks in $\Lagr_i$ can be seen. 
These can be accounted to LBFGS algorithm restarts, since the Hessian approximation is bad initially and improves after several iterations of LBFGS.
One can still see that the final values of the LBFGS iterations before the restarts have lower $\Lagr_i$ loss values than the previous LBFGS iterations.
That means the restarts improve the approximation accuracy of the displacement field. 
In Fig.~\ref{Fig:conv_outer_3d} the convergence of the dual update is shown.
The curve on the figure indicates a fast convergence of the criterion within $10$ iterations. 
Here, the slope at the last iteration is not zero. 
The dual update is stopped since tolerance $\varepsilon^{\li}_{tol}$ is reached.

\begin{figure}[b!]
    \centering

    \begin{subfigure}[t]{0.3\textwidth}
        \centering
        \includegraphics[width=\linewidth]{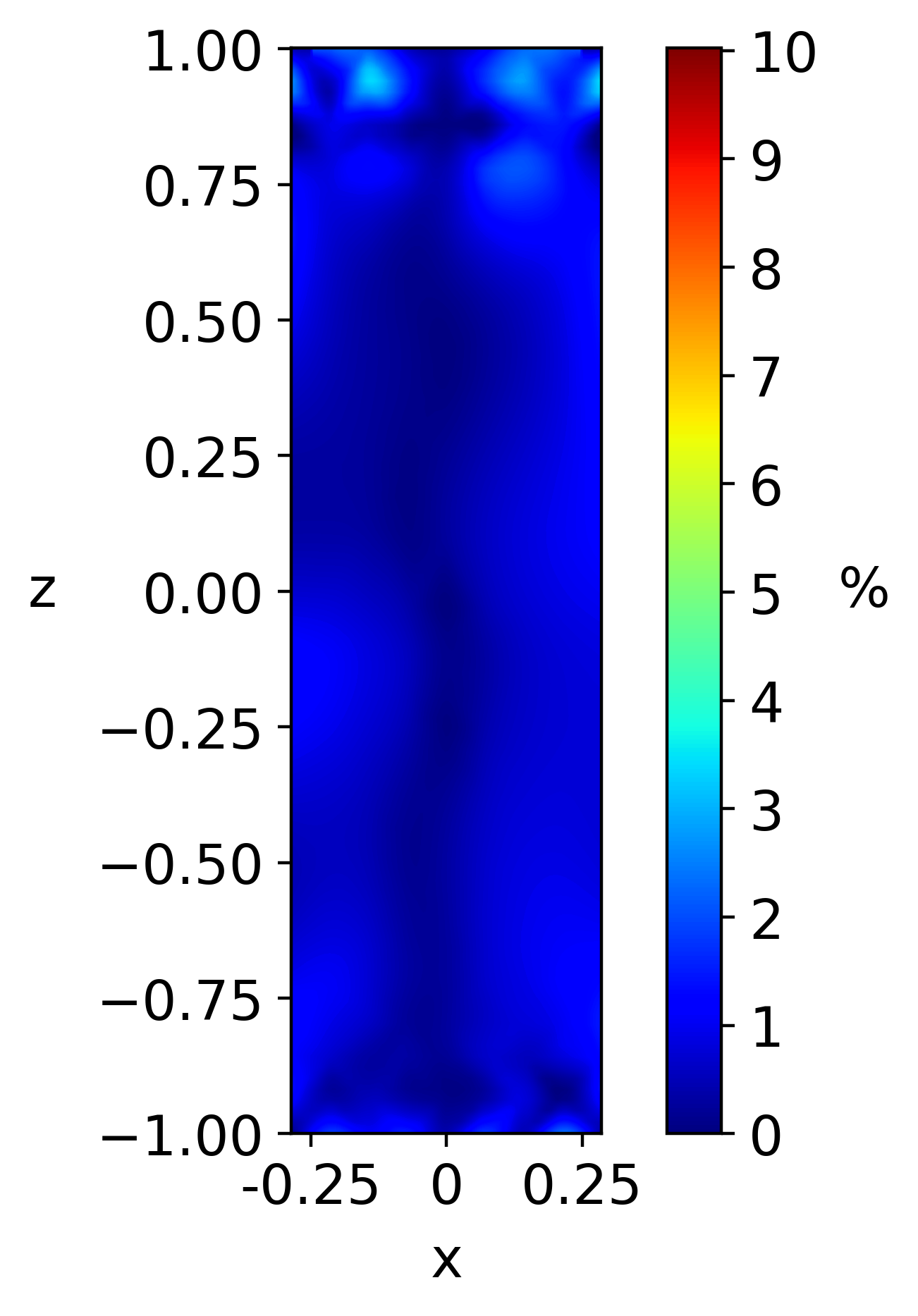}
        \caption{Mean relative error $u_x$.}
    \end{subfigure}
    \hfill
    \begin{subfigure}[t]{0.3\textwidth}
        \centering
        \includegraphics[width=\linewidth]{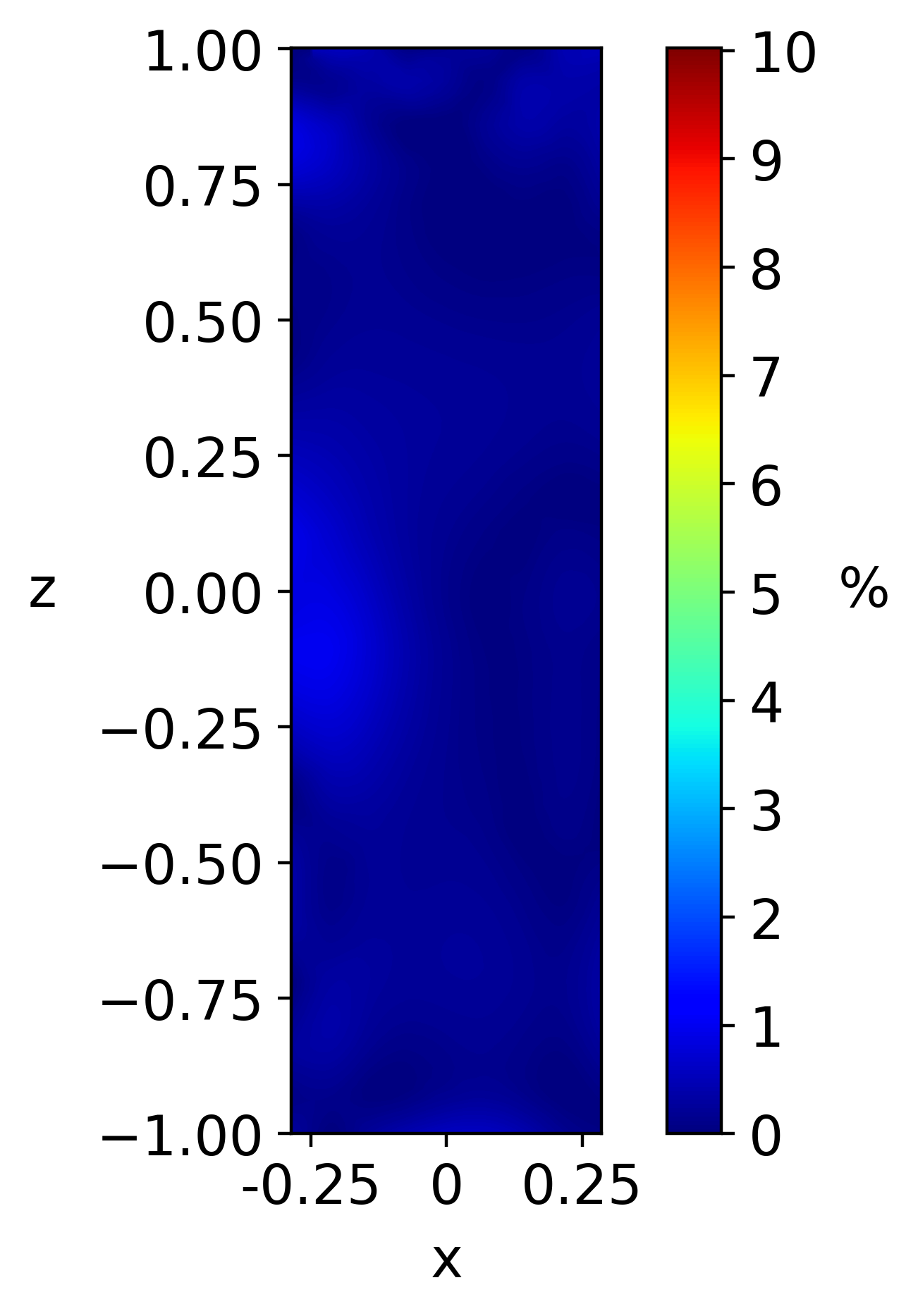}
        \caption{Mean relative error $u_y$.}
    \end{subfigure}
    \hfill
    \begin{subfigure}[t]{0.3\textwidth}
        \centering
        \includegraphics[width=\linewidth]{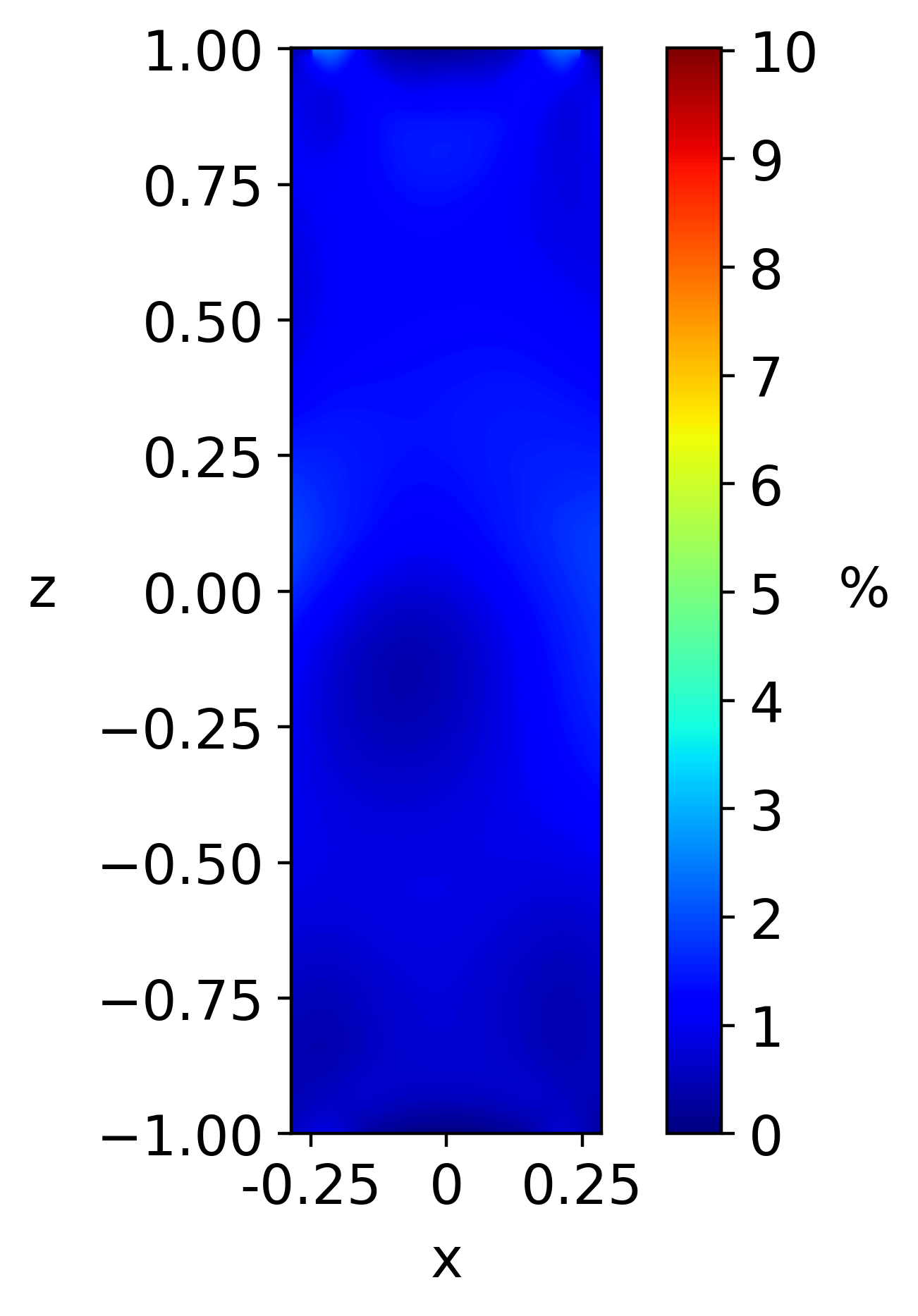}
        \caption{Mean relative error $u_z$.}
    \end{subfigure}

    \vspace{1em} 

    \begin{subfigure}[t]{0.3\textwidth}
        \centering
        \includegraphics[width=\linewidth]{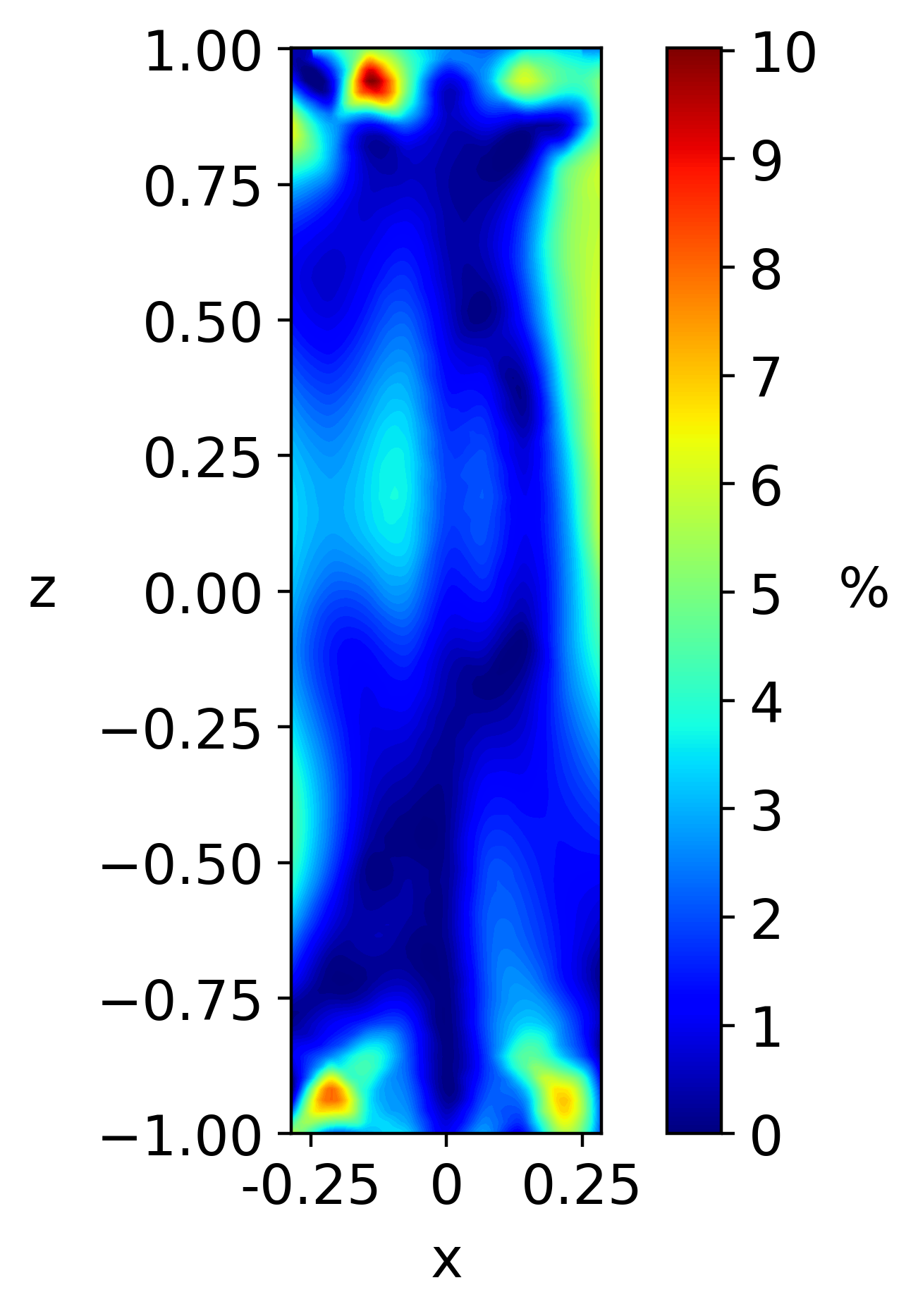}
        \caption{Standard deviation relative error $u_x$.}
    \end{subfigure}
    \hfill
    \begin{subfigure}[t]{0.3\textwidth}
        \centering
        \includegraphics[width=\linewidth]{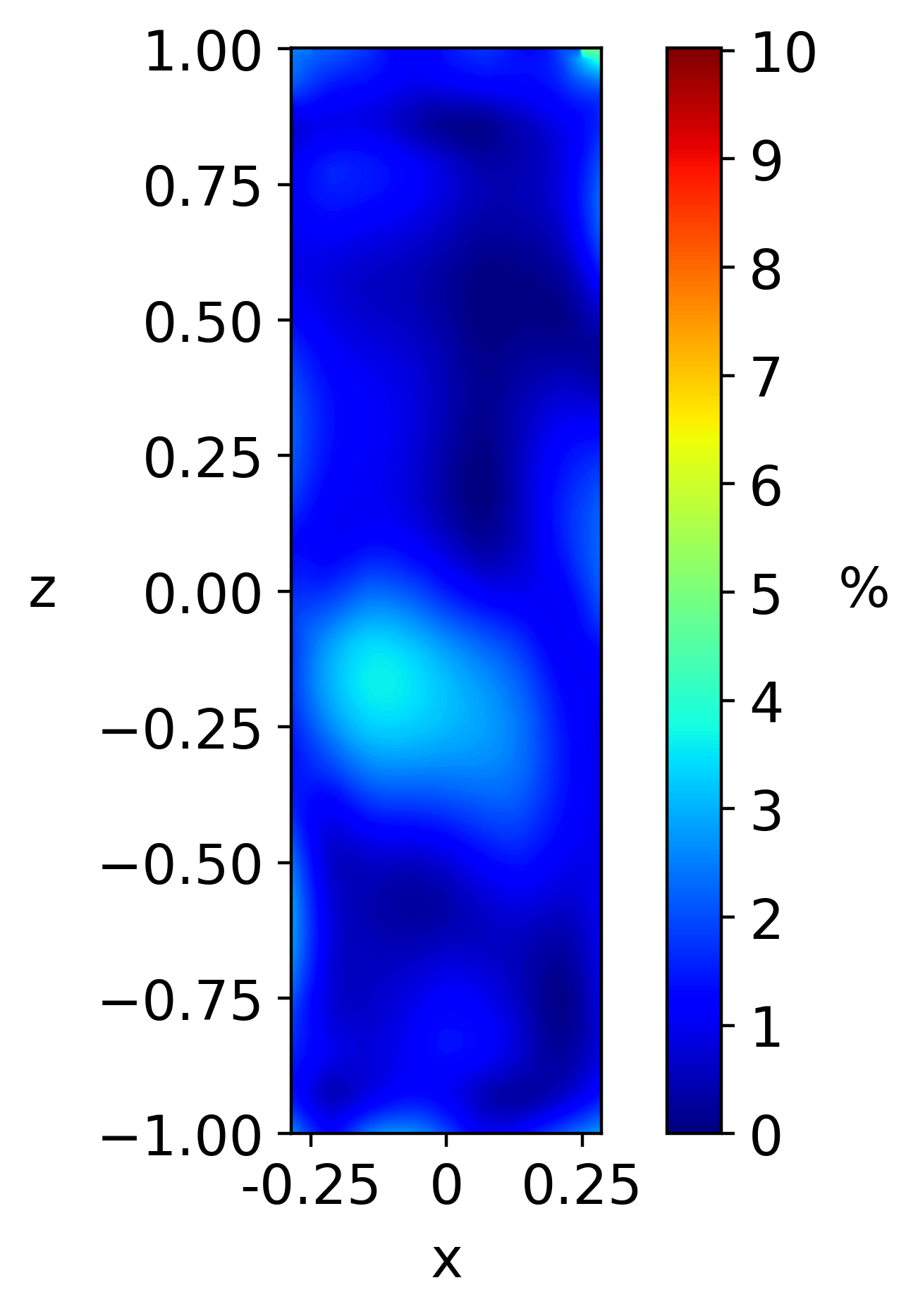}
        \caption{Standard deviation relative error $u_y$.}
    \end{subfigure}
    \hfill
    \begin{subfigure}[t]{0.3\textwidth}
        \centering
        \includegraphics[width=\linewidth]{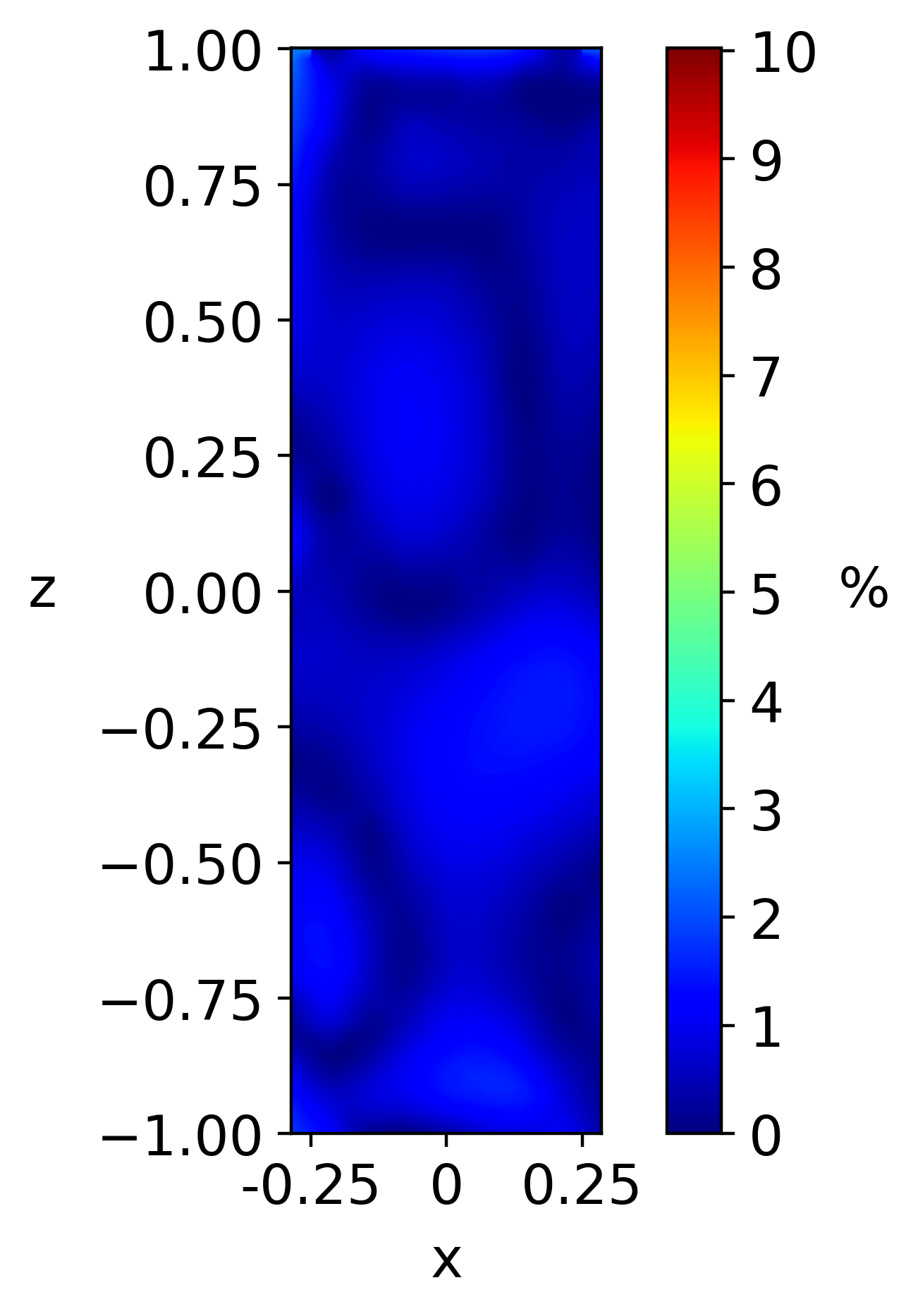}
        \caption{Standard deviation relative error $u_z$.}
    \end{subfigure}

    \caption{Interpolation plots of the relative errors in mean and standard deviation of the displacement components $\fvar$ approximated by the NNs compared to data on a section through the cylinder. Values are normalized to allow optical comparison to the two-dimensional problem.}
    \label{Fig:ints_ms}
\end{figure}

The accuracy of the approximation is assessed by comparing the model and data statistics by computing the mean and standard deviation across the entire domain. 
The mean and standard deviations are:
\begin{equation*}
    \mathbb{E}(u(x)) \approx \frac{1}{N_u} \sum_{i=1}^{N_u} u_i(x),
\end{equation*}
and
\begin{equation*}
    \sqrt{var(u(x))} \approx \sqrt{\frac{1}{N_u} \sum_{i=1}^{N_u} \left( u_i(x) - \frac{1}{N_u} \sum_{j=1}^{N_u} u_j(x) \right)^2},
\end{equation*}
computed over the 1000 training points and also over 1000 test points, which are indicated as $N_u$.
To visualize the quality of the fit, Fig.~\ref{Fig:ints_ms} presents results of these statistics on a cross-section at $y = 0$ for the training points.
The spatial and displacement components in the figure are normalized to allow a visual comparison of scale of the errors between the example in the first section and the example in this section.
The figure shows the relative error, defined in Eq. \ref{Eq:rele+1}, for both the mean and standard deviation over the resulting $x$–$z$ plane.
In the first row, the mean relative errors are shows for the three displacement components, while the second row shows the corresponding standard deviation relative errors. 
\begin{figure}[!ht]
    \centering

    \begin{subfigure}[t]{0.3\textwidth}
        \centering
        \includegraphics[width=1.15\linewidth]{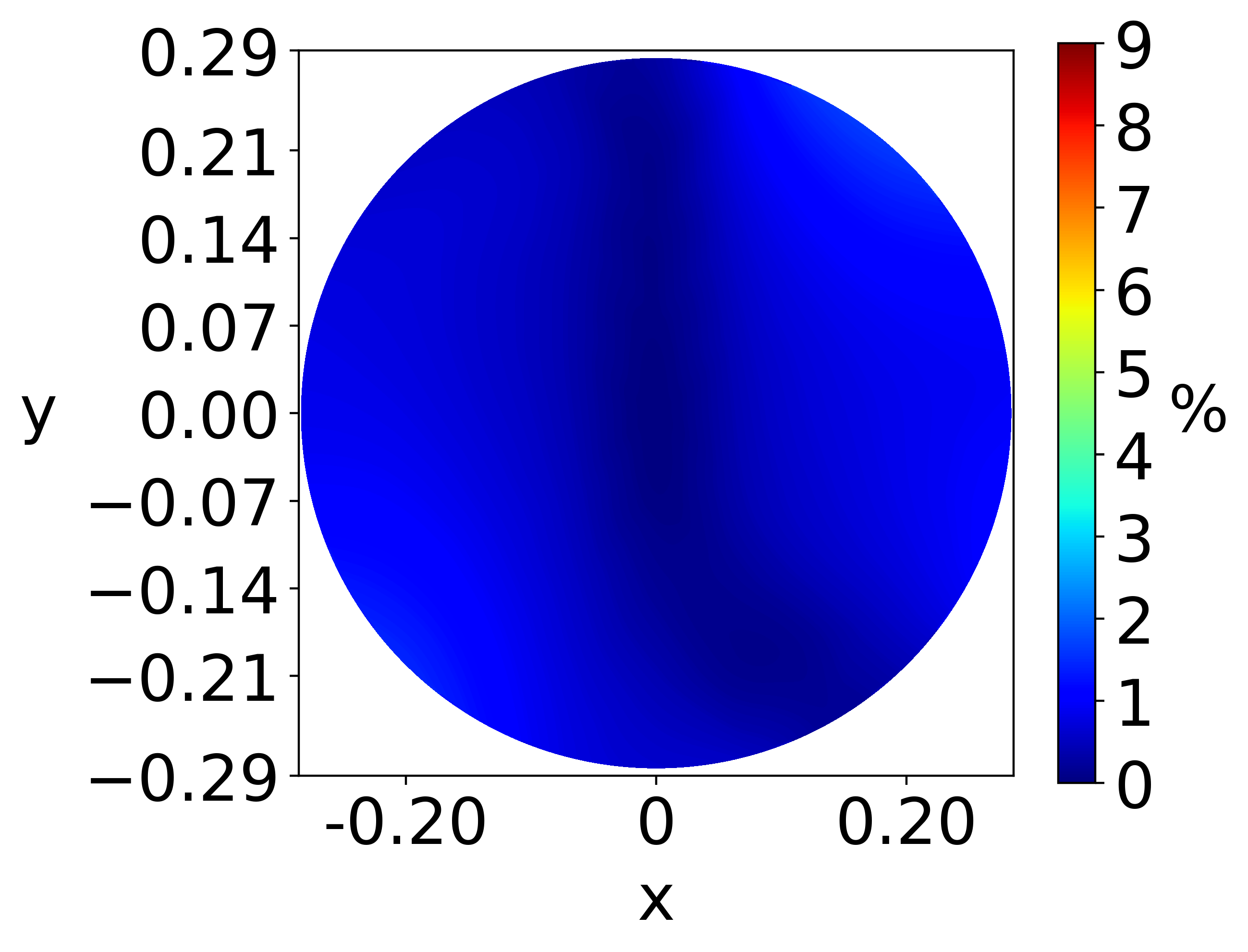}
        \caption{Mean relative error $u_x$.}
    \end{subfigure}
    \hfill
    \begin{subfigure}[t]{0.3\textwidth}
        \centering
        \includegraphics[width=1.15\linewidth]{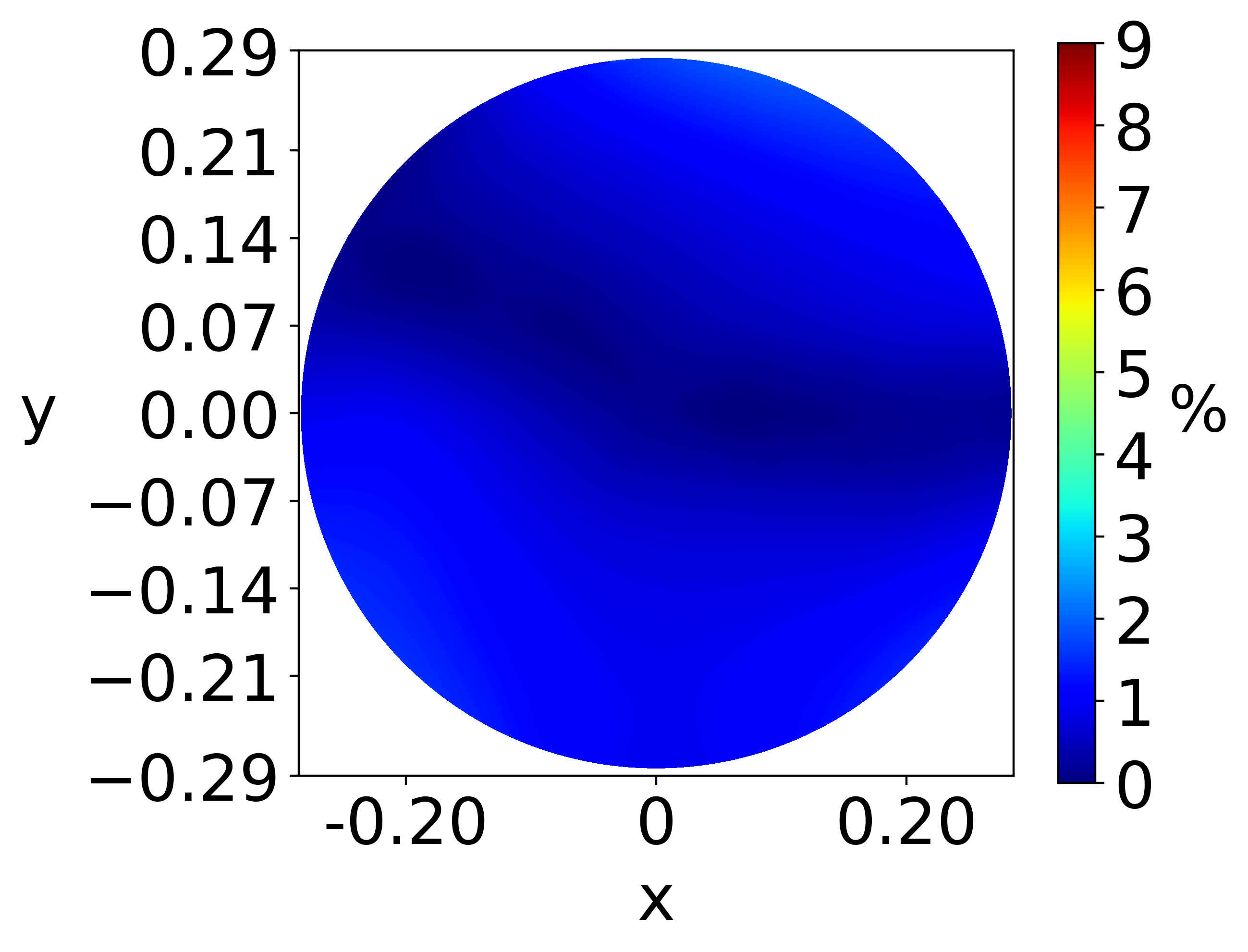}
        \caption{Mean relative error $u_y$.}
    \end{subfigure}
    \hfill
    \begin{subfigure}[t]{0.3\textwidth}
        \centering
        \includegraphics[width=1.15\linewidth]{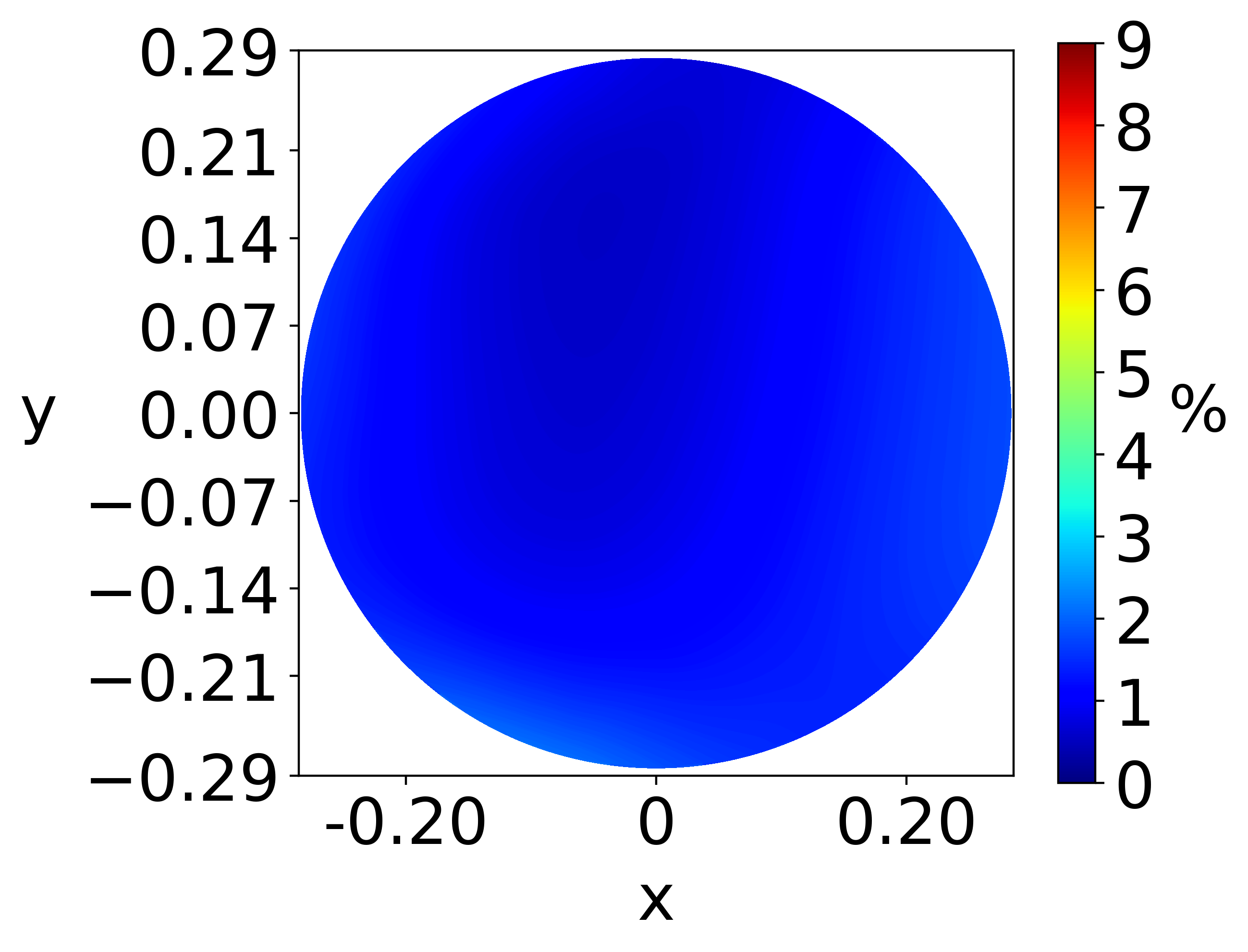}
        \caption{Mean relative error $u_z$.}
    \end{subfigure}

    \vspace{1em} 

    \begin{subfigure}[t]{0.3\textwidth}
        \centering
        \includegraphics[width=1.15\linewidth]{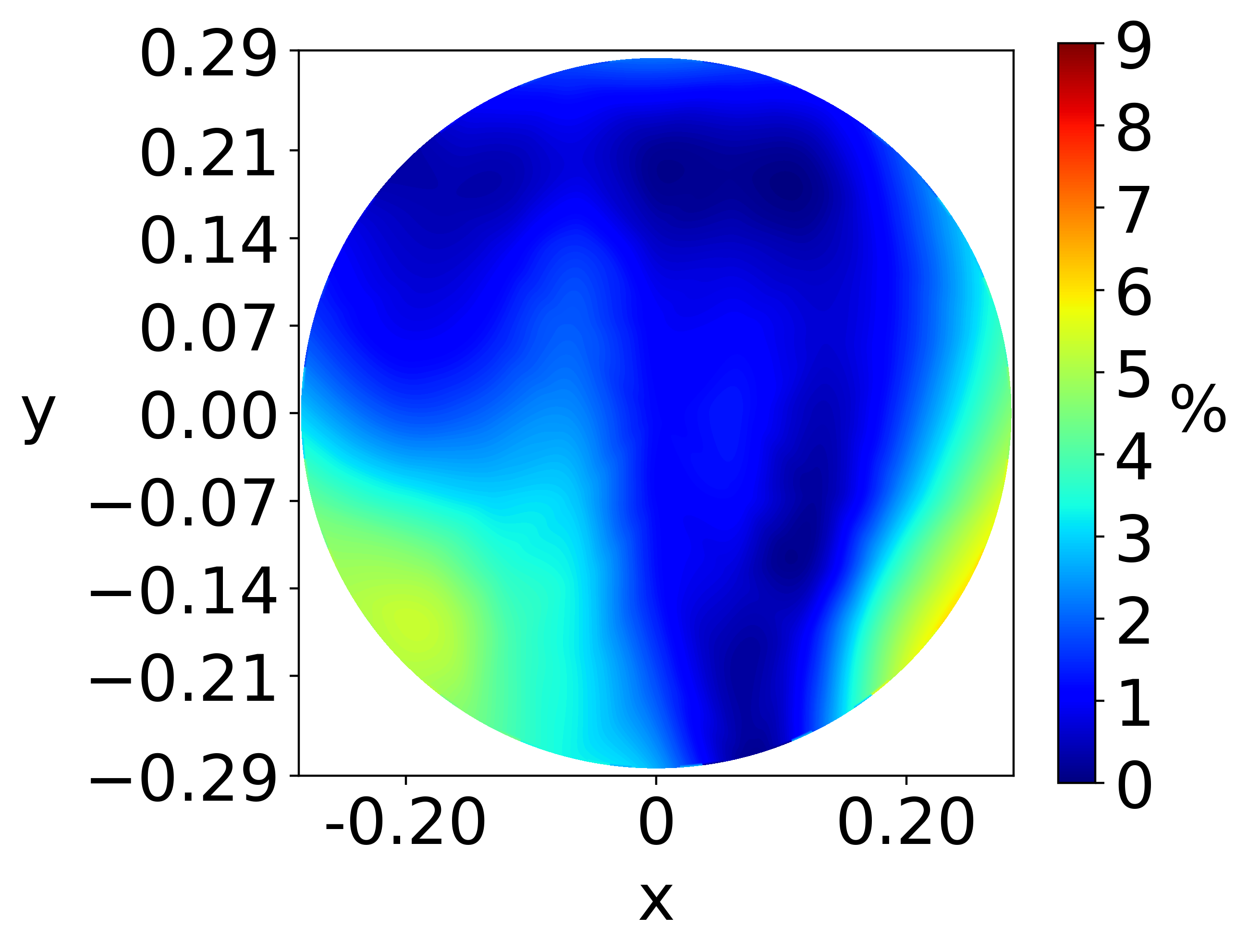}
        \caption{Standard deviation relative error $u_x$.}
    \end{subfigure}
    \hfill
    \begin{subfigure}[t]{0.3\textwidth}
        \centering
        \includegraphics[width=1.15\linewidth]{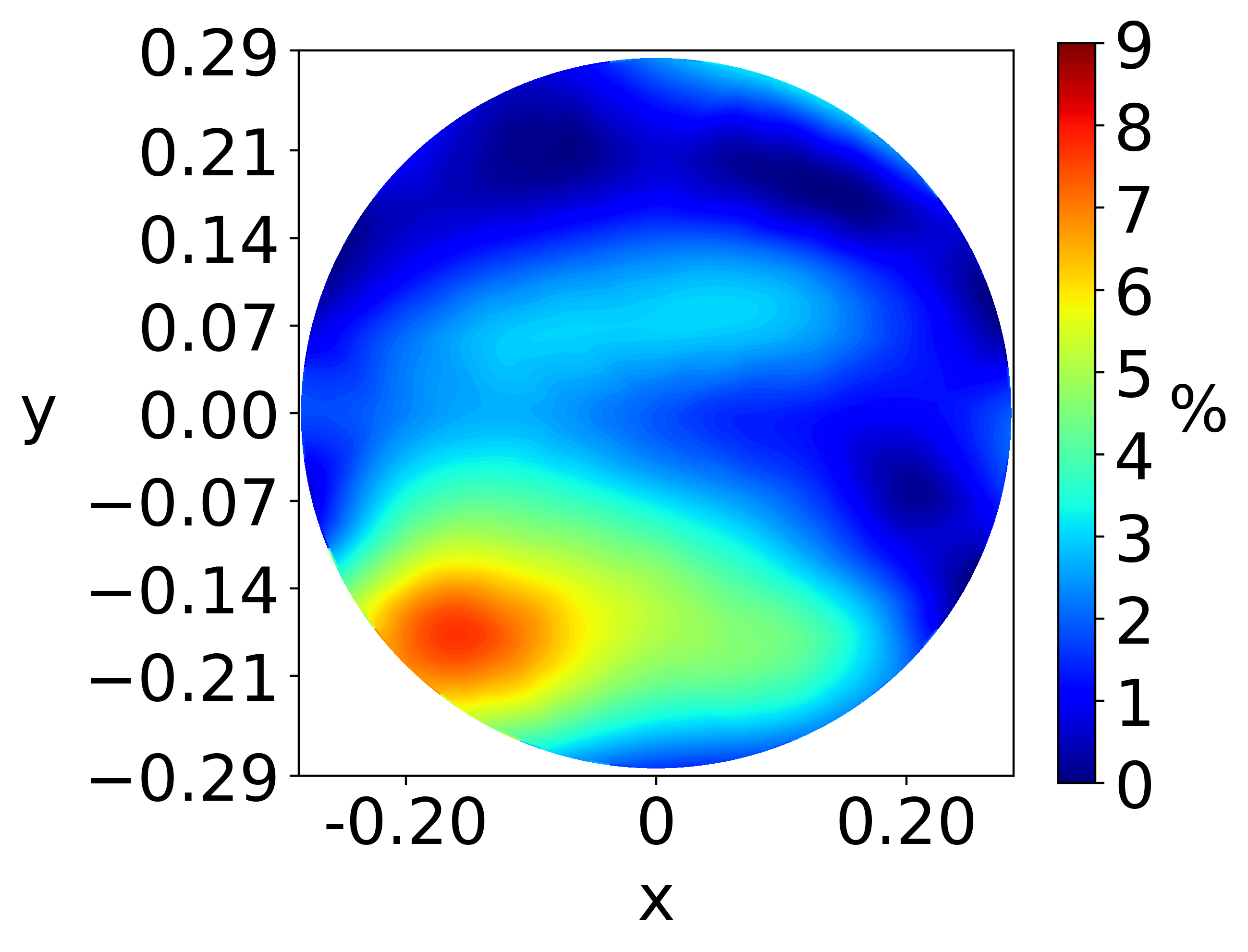}
        \caption{Standard deviation relative error $u_y$.}
    \end{subfigure}
    \hfill
    \begin{subfigure}[t]{0.3\textwidth}
        \centering
        \includegraphics[width=1.15\linewidth]{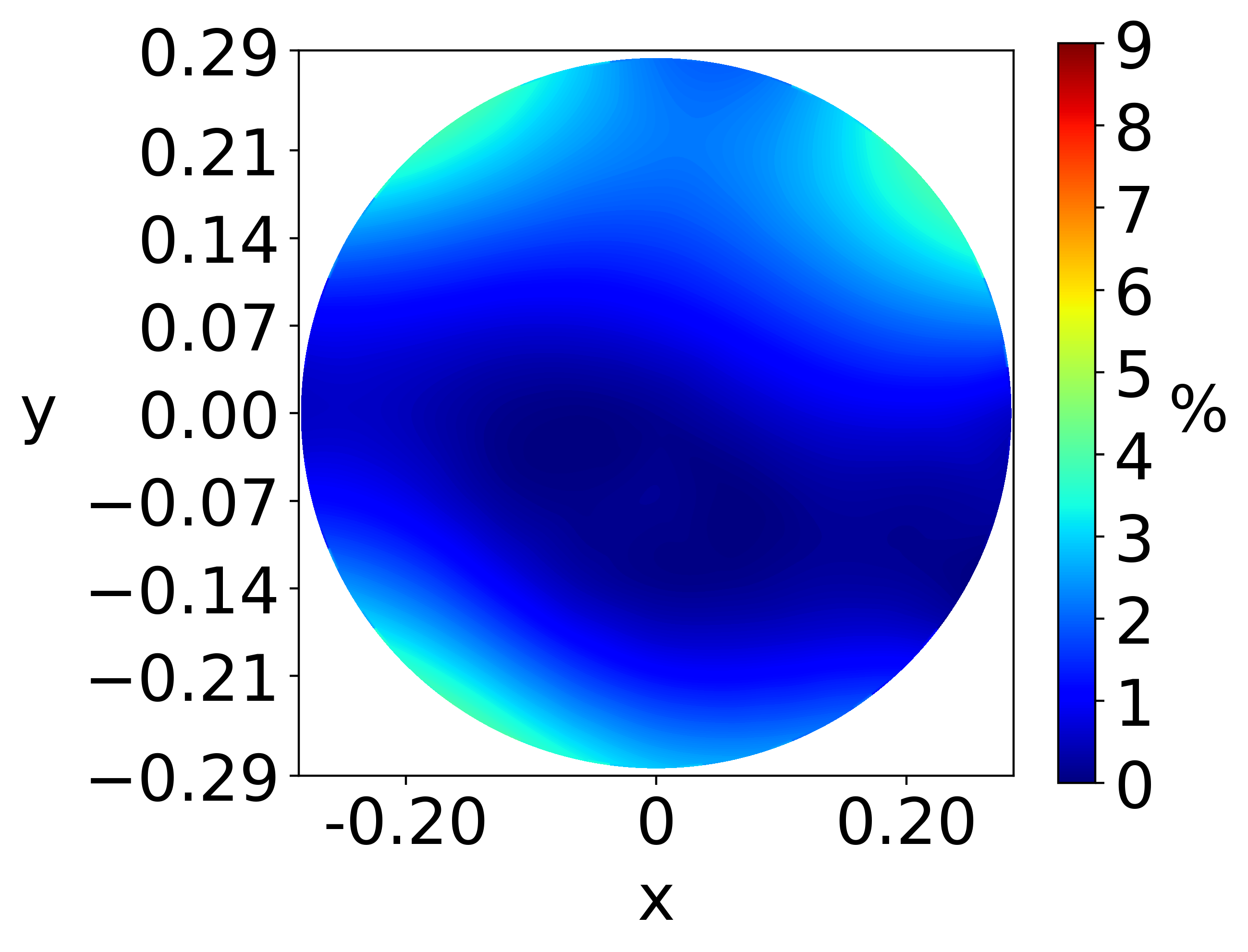}
        \caption{Standard deviation relative error $u_z$.}
    \end{subfigure}

    \caption{Interpolation plots of the relative errors in mean and standard deviation of the displacement components $\fvar$ approximated by the NNs compared to data on the interface. Values are normalized to allow optical comparison to the two-dimensional problem.}
    \label{Fig:ints_msi}
\end{figure}
It can be seen that the mean displacement values across the domain inherit a negligible error, with the maximum error in the $u_x$ component around $2.5\%$ at the top of the domain.
All standard deviations show slightly larger relative errors since they depend on second-order moments, which are more sample-sensitive.
The standard deviation of the $u_x$ component shows the largest relative error of $10\%$ at the same point as the mean.
Results reveal that the relative error is small and could be readily reduced by increasing the number of samples.
This indicates that the surrogate model accurately approximates the displacement field across the range of random material parameters.
Also, no noticeable deviations are observed along the interface at $0$, which is shown in more detail in Fig.~\ref{Fig:ints_msi}.
The figure shows the mean and standard deviation relative error components for each displacement component at the interface for training points. 
In the first row, the mean components are shown and in the second row the standard deviations are shown. 
As is previously found, the mean relative error is very low at the interface, while some relative errors in standard deviation exist.
The standard deviation errors are larger near the domain boundaries.
Larger errors near the domain boundary can be accounted to the smaller number of constraints around an edge point. 
Internal points are instead constrained also by points all around.
The highest standard deviation error on the interface is $8\%$ in the $u_y$ component, while a large part of the interface has errors below $3\%$.
For the given number of samples, the results indicate that the \al worked and the constraints are satisfied.
Further improvements can be achieved by increasing the number of training samples.

\begin{figure}[!ht]
    \centering 
    \includegraphics[width=0.6\textwidth]{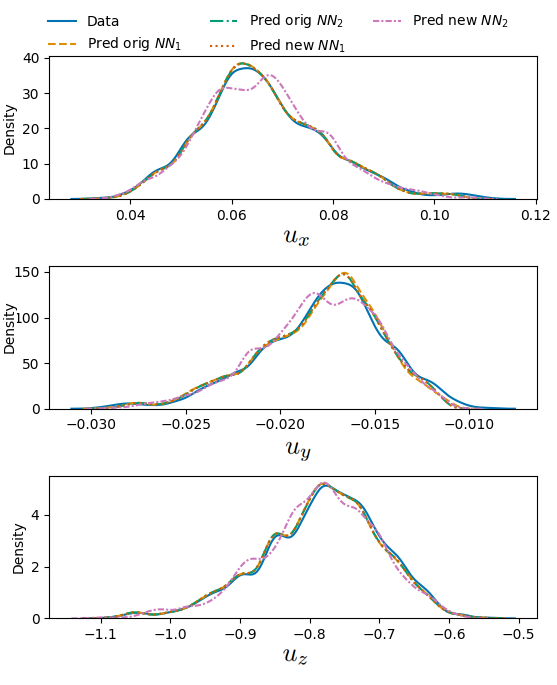}
    \caption{Kernel density estimator distributions of an interface point of the three-dimensional problem for each displacement component.} 
    \label{Fig:p_distr}
\end{figure}

To assess the accuracy at the local level, the displacement distribution at a single point located on the interface between the two subdomains is examined. 
This analysis compares the data distribution with the prediction distributions of each subdomain for both training and test samples. 
Fig.~\ref{Fig:p_distr} presents the distributions as kernel density estimates, with displacement components on the horizontal axis and probability densities on the vertical axis. 
The blue curves show the FEM sample distributions, which are very similar compared to the approximated distributions of both neural networks, indicating a high continuity across the interface. 
Predictions of $NN_1$ from training (green) and test samples (yellow) show slightly higher peaks than the sampling distributions, though the difference is small. 
The lower-domain network ($NN_1$) reproduces the reference distribution slightly more accurately than the upper-domain network ($NN_2$) at the interface. 
However, predictions from both subdomains remain close to the reference data. 
These results demonstrate that the surrogate model provides sufficiently accurate predictions for use in optimization.
It should be noted, that the optimization parameters, due to the increase in dimensionality, lead to a substantial growth in the number of Lagrange multipliers. 
Consequently, the computational cost rises, and managing the large set of multipliers becomes challenging. 
Future work should investigate alternative approaches, to reduce the number of interface parameters while maintaining continuity.

\section{Conclusion}
\label{Sec:final}
Two algorithms are proposed to decompose global NN approximations into simpler, local models. 
Each local NN is independently optimized using a local loss function and the L-BFGS optimization algorithm.
Interface continuity between local NN domains is enforced through Lagrange multipliers, introduced via modifications to the local loss functions. 
In the augmented Lagrange formulation, an additional penalty term is included to regularize the constraints.

Both algorithms are evaluated on a two-dimensional linear-elastic compression problem, using data generated from the FEM. 
The goal is to approximate the mapping from spatial input coordinates to the corresponding horizontal displacements. 
Multiple spatial decompositions are tested to identify an optimal configuration with three vertically stacked subdomains showing the best performance.
Three vertically stacked subdomains took $10$ h  to compute and have a maximum relative error of $4\%$ for \al across the whole domain. 

In all tested decompositions, both methods significantly reduce discontinuities at the interfaces between subdomains, demonstrating their effectiveness in enforcing continuity.
However, \al converges to a precise approximation faster when the penalty parameter is correctly chosen. 
In contrast, \nl often struggles to identify a feasible linearization during updates, frequently requiring multiple attempts before convergence.
This limitation makes \nl less suitable for highly nonlinear problems, where the linearization around the current weights may not adequately capture the true behavior of the objective.
For highly nonlinear functions, improved linearization strategies or more computationally demanding approaches, e.g., second-order approximations, may be required to achieve convergence. 
Overall, \al shows greater reliability and scalability, making it more suitable for training large NNs.

Due to the high cost and a limited number of sensors, measurement data is often sparse. 
Consequently, the NN DDMs can be used to bridge between data-rich regions.
Therefore, the impact of gaps at the interfaces and number of points per subdomain on the performance of the \al is investigated.
Stability of the method was observed across a broad range of numbers of points per domain and interface gap sizes between domains.
However, a gap between domains results in an uncertainty propagation problem. 
Hence, the approximation will converge to a different solution every time.
\al is a powerful method to parallelize the training process of NNs for large solution domains with local nonlinearities. 

In the future, \al will be applied to larger-scale problems to further assess its scalability in comparison to standard, un-decomposed NNs. 
While the present study primarily demonstrates improvements in approximation accuracy and introduces the method, larger problems would highlight also the method’s computational advantages. 
Moreover, such problems would highlight the benefits of parameter reduction within local domains. 
Additionally, the gaps between local domains could be modeled as random fields or NNs.
This will allow convergence toward a single, stochastic approximation on the interfaces and lower the increased computational cost of additional dimensions through samples of random variables. 
\section{Acknowledgements}
This research was carried out under project number T21001a in the framework of the Partnership Program of the Materials innovation institute M2i (www.m2i.nl).

\printbibliography
\end{document}